\documentclass{aastex631}
\usepackage{makecell}
\usepackage{amsmath}
\usepackage{cases}
\usepackage{multirow}

\begin{document}

\title{Revision of the GeV $\gamma$-ray Emission in the Region of HESS J1813-178 with Fermi-LAT}
\author[0009-0004-5450-8237]{Xiaolei Guo}
\author[0000-0001-5135-5942]{Yuliang Xin}
\affiliation{School of Physical Science and Technology, Southwest Jiaotong University, Chengdu 610031, People's Republic of China;
\href{mailto:xlguo@swjtu.edu.cn}{xlguo@swjtu.edu.cn}, \href{mailto:ylxin@swjtu.edu.cn}{ylxin@swjtu.edu.cn}}



\begin{abstract}
HESS J1813-178 is one of the brightest and most compact TeV $\gamma$-ray sources, and whether its $\gamma$-ray emission is associated with supernova remnant (SNR), pulsar wind nebula (PWN) or young stellar cluster (YSC) is still under debate. By analysing the GeV $\gamma$-ray data in the field of HESS J1813-178 using 14 years of PASS 8 data recorded by the {\em Fermi} Large Area Telescope ({\em Fermi}-LAT), we report the discovery of three different sources with different spectra in this region. The hard source with a power law spectral index of 2.11 $\pm$ 0.08 has a small size extension, which is spatially and spectrally coincident with the TeV $\gamma$-ray emission from HESS J1813-178. 
CO observations display the dense molecular clouds surrounding HESS J1813-178 in the velocity range of 45-60 km s$^{\rm -1}$.
The possible origins of the $\gamma$-ray emission from HESS J1813-178 are discussed, including SNR G12.82-0.02, the PWN driven by the energetic X-ray pulsar PSR J1813-1749, and YSC Cl 1813-178.
However, none of them can be ruled out clearly.
%The incompatibility between the hard X-ray spectrum of PWN and the flat GeV $\gamma$-ray spectrum of HESS J1813-178 suggests a two-zone leptonic model for the PWN scenario. The $\gamma$-ray spectrum of HESS J1813-178 can also be well fitted by a hadronic model, in which the proton could be accelerated and escaped from SNR G12.82–0.02 in earlier evolutional epochs. Although the flat $\gamma$-ray spectrum of HESS J1813-178 is similar to that of YSCs with the typical spectral index of 2.1-2.3, the GeV extension of it is much smaller than that of other YSCs, which suggests a low diffusion coefficient being close to Bohm limit.
%Although either the hadronic emission associated with SNR G12.82–0.02 or YSC Cl 1813-178 can match the $\gamma$-ray observations, the hard GeV $\gamma$-ray spectrum of HESS J1813-178 is much distinguished from that of the typical old-aged SNRs interacting with molecular clouds, but is similar to that of YSCs with the typical spectral index of 2.1-2.3.
Note that the maximum energy of protons in the hadronic model should exceed a few hundred TeV, which makes HESS J1813-178 to be a promising PeVatron. The detailed LHAASO data analysis about the morphology and spectrum would be helpful to investigate the origin of the $\gamma$-ray emission in this region and test its PeVatron nature.
\end{abstract}



\section{Introduction}
\label{intro}

Very high energy (VHE; $>$100 GeV) surveys of our Galaxy with the ground based $\gamma$-ray detectors, like H.E.S.S., MAGIC, HAWC and LHAASO, have uncovered a population of $\gamma$-ray sources in the TeV regime and revealed the very energetic cosmic accelerators in our Galaxy. 
More than two hundred of VHE $\gamma$-ray sources have been detected\citep{2018A&A...612A...1H,2020ApJ...905...76A,2023arXiv230517030C}, most of which have been identified to be PWNe, SNRs, binaries, TeV halos and so on. 
However, a large fraction of the sources still remain without firm association, and the multi-wavelength observations of these sources are crucial for revealing their nature and elucidate the origin of cosmic rays (CRs) around the knee energy range.

As one of the brightest and most compact objects detected by the H.E.S.S. Galactic Plane Survey (HGPS) observations, HESS J1813-178 was detected to be nearly point-like with an extension of $\sigma=2.'2$ \citep{2005Sci...307.1938A, 2006ApJ...636..777A}. 
The TeV $\gamma$-ray emission from HESS J1813-178 shows a power-law spectrum with a rather hard photon index of 2.09$\pm$0.08, and the following MAGIC observations give the similar spectrum with an index of 2.1$\pm$0.2 in the energy range of 0.4-10 TeV \citep{2006ApJ...637L..41A}.
In \citet{2018A&A...612A...1H}, the extension of HESS J1813-178 was updated to be $\sigma=3.'0 \pm 0'.24$, and the spectrum above 1 TeV follows a power law with an exponential cutoff of $\sim$ 7 TeV.

In the First LHAASO Catalog of Gamma-Ray Sources, both the Water Cherenkov Detector Array (WCDA) and Kilometer Squared Array (KM2A) of LHAASO detected a $\gamma$-ray source in the region of HESS J1813-178, 1LHAASO J1814-1719u, in the energy range of 1-25 TeV and above 25 TeV, respectively \citep{2023arXiv230517030C}.
The spatial template of 1LHAASO J1814-1719u for WCDA is described by a two dimensional (2D) Gaussian with $\sigma = 0.71^\circ \pm 0.07^\circ$, and its $\gamma$-ray spectrum in 1-25 TeV is modeled by a power law with an index of 2.83 $\pm$ 0.06.
KM2A gives an upper limit of the extension of 1LHAASO J1814-1719u with $\sigma = 0.27^\circ$, together with the power law spectral index of 3.49 $\pm$ 0.31 above 25 TeV \citep{2023arXiv230517030C}.

HESS J1813-178 is positionally coincident with a shell-type SNR G12.82-0.02, which lies in the vicinity of a bright star forming region (SFR) W33 \citep{2005ApJ...629L.105B}.
And non-thermal X-ray emission was also detected in this region \citep{2005ApJ...629L.105B, 2006ApJ...651..190L}.
Deep X-ray observations from {\em XMM}-Newton and {\em Chandra} revealed a PWN embedded near the center of SNR G12.82-0.02, which makes G12.82-0.02 to be a composite system \citep{2007A&A...470..249F, 2007ApJ...665.1297H, 2009ApJ...700L.158G}.
The PWN component is powered by an energetic pulsar, PSR J1813-1749, and its spin-down luminosity and characteristic age are $\dot{E} = 5.6 \times 10^{37} \ \rm erg \ s^{-1}$ and $\tau_c = 5600 \rm \ yr$, respectively  \citep{2012ApJ...753L..14H, 2020MNRAS.498.4396H}.
A young stellar cluster at a kinematic distance of 4.8 kpc, Cl 1813-178, was discovered in the TeV $\gamma$-ray emission region of HESS J1813-178 \citep{2008ApJ...683L.155M, 2011ApJ...733...41M}.
PSR J1813-1749/SNR G12.82-0.02 was first suggested to be associated with Cl 1813-178 at the same distance \citep{2011ApJ...733...41M}.
While the newest radio observation from Green Bank Telescope shows a large dispersion measure for PSR J1813-1749, which indicates a far distance of either 6.2 or 12 kpc according to the different models of the electron density distribution \citep{2021ApJ...917...67C}.

The origin of the $\gamma$-ray emission from HESS J1813-178 is still not identified.
The scenarios of PWN, SNR and stellar cluster are discussed in previous works.
Using the NANTEN $^{\rm 12}$CO($J$=1-0) observations, \citet{2007A&A...470..249F} detected a giant molecular cloud (MC) in the south of HESS J1813-178, which is suggest to be associated with SFR W33. 
With the X-ray and molecular observations, the scenarios in which the $\gamma$-rays of HESS J1813-178 are emitted from the shell of the SNR and one in which they are emitted from the central PWN are modeled. None of them can be clearly ruled out.
\citet{2010ApJ...718..467F} also predicted that the TeV $\gamma$-ray emission is mainly originated from the PWN associated with PSR J1813-1749, although the contribution from the SNR shell could be enhanced with a denser medium.
With the {\em Fermi}-LAT data, \citet{2018ApJ...859...69A} found extended GeV $\gamma$-ray emission around HESS J1813-178 with a radius of $\sim$0.6$^{\circ}$.
Such size is much larger than that of the TeV $\gamma$-ray emission of HESS J1813-178.
The global GeV spectrum in the energy range of 0.5-500 GeV was fitted by a power-law with an index of 2.14$\pm$0.04, which is not consistent with the inverse Compton scattering (ICS) emission characteristic from leptons in a PWN. 
\citet{2018ApJ...859...69A} argued that the extended GeV emission may be related to the SFRs around HESS J1813-178, like W33.

Considering the uncertain origin and the complexity of HESS J1813-178, the detailed analysis with more GeV observational data and the multi-wavelength observations would be helpful to investigate the origin of the $\gamma$-ray emission in this region, we re-analysed the {\em Fermi}-LAT PASS 8 data around HESS J1813-178 in this work.
The data analysis procedure and results, including the spatial and spectral analysis, are shown in Section 2.
In Section 3, the observation of the molecular clouds around HESS J1813-178 is introduced. 
We discuss the origin of the $\gamma$-ray emission from HESS J1813-178 based on the multi-wavelength observations in Section 4.
The conclusion of this work is presented in Section 5.

\section{{\em Fermi}-LAT Data Analysis}
\label{fermidata}

\subsection{Data Reduction}
\label{data reduction}

In the following analysis, the latest Pass 8 data with ``Source'' event class are selected 
from 2008 August 4 (MET 239557418) to 2022 August 4 (MET 681264005).
The energies of events are cut between 500 MeV and 1 TeV to avoid a too large point spread function (PSF) in the lower energy band.
Meanwhile, the maximum zenith angle is adopted to be 90$^\circ$ to minimize the contamination from the Earth Limb.
The analysis is performed in a 14$^\circ \times 14^\circ$ square region of interest (ROI) with the standard LAT analysis software {\it ScienceTools}.
The binned likelihood analysis method with {\em gtlike} is adopted, together with the instrumental response function of ``P8R3{\_}SOURCE{\_}V3''.
The Galactic and isotropic diffuse backgrounds are modeled by 
{\tt gll\_iem\_v07.fits} and {\tt iso\_P8R3\_SOURCE\_V3\_v1.txt}, respectively.
All sources listed in the incremental version of the fourth {\em Fermi}-LAT source catalog \citep[4FGL-DR4;][]{2022ApJS..260...53A,2023arXiv230712546B} within a radius of 20$^\circ$ from the ROI center, together with the two diffuse backgrounds, are included in the model, which is generated with the user-contributed script {\tt make4FGLxml.py}.

\subsection{Spatial Analysis}
\label{spatial}

\begin{figure*}[!htb]
	\centering
    \includegraphics[width=0.5\textwidth]{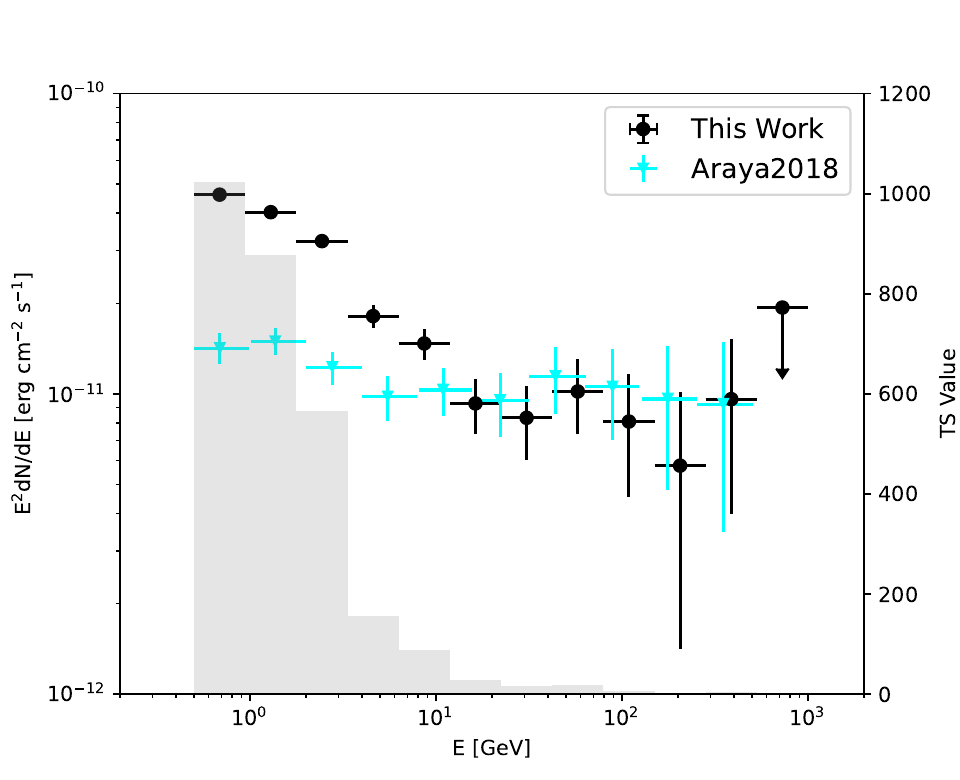}
	\caption{The SED of 4FGL J1813.1-1737e with black dots. The arrows indicate the 95\% upper limits, and the gray histogram denotes the TS value for each energy bin. The cyan dots are the results from \citet{2018ApJ...859...69A}.}
	\label{fig1:sed}
\end{figure*}

In 4FGL-DR4 catalog, there is an extended GeV $\gamma$-ray source (4FGL J1813.1-1737e) described by an uniform disk with a radius of 0.6$^\circ$, which is regarded as the potential GeV counterpart of HESS J1813-178 \citep{2018ApJ...859...69A, 2022ApJS..260...53A}.
However, the size of 4FGL J1813.1-1737e is much larger than that of TeV $\gamma$-ray emission from HESS J1813-178 \citep[$\sigma=3.'0$;][]{2018A&A...612A...1H}.
To investigate the GeV $\gamma$-ray emission around HESS J1813-178, 
we first refit the centroid and size of the uniform disk for 4FGL J1813.1-1737e with {\tt Fermipy}, a {\tt PYTHON} package that automates analyses with {\it ScienceTools} \citep{2017ICRC...35..824W}.
The best-fit centroid position and 68\% containment radius (R$_{\rm 68}$) are fitted to be R.A. = $273^{\circ}\!.419 \pm 0^{\circ}\!.019$, decl. = $-17^{\circ}\!.625 \pm 0^{\circ}\!.018$, and $0^{\circ}\!.475 ^{+0^{\circ}\!.015}_{-0^{\circ}\!.014}$, respectively.
With the refit spatial model (Model 1), we get its GeV spectrum by dividing the data into 12 equal logarithmic energy bins from 500 MeV to 1 TeV. 
The likelihood fitting analysis is done for each energy bin and the 95\% upper limits are calculated for the energy bins with Test Statistic \citep[TS;][]{2012ApJ...756....5L} values smaller than 5.0.
The resulting spectral energy distribution (SED) of 4FGL J1813.1-1737e is shown in Figure \ref{fig1:sed}, and an obvious spectral upturn is shown below an energy of about 10 GeV compared with the results from \citet{2018ApJ...859...69A}.

\begin{figure*}[!htb]
	\centering
	\includegraphics[width=0.49\textwidth]{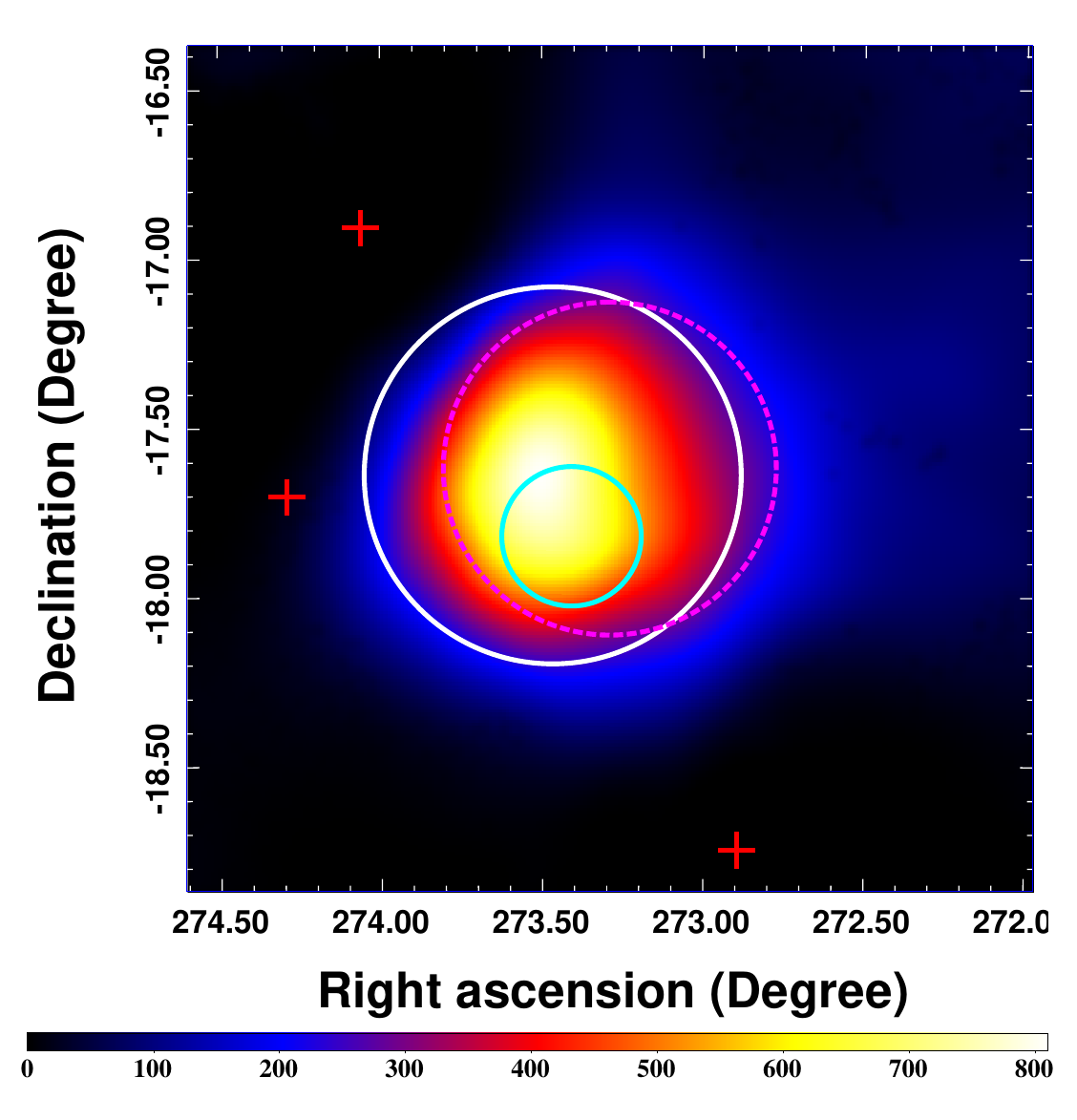}
        \includegraphics[width=0.49\textwidth]{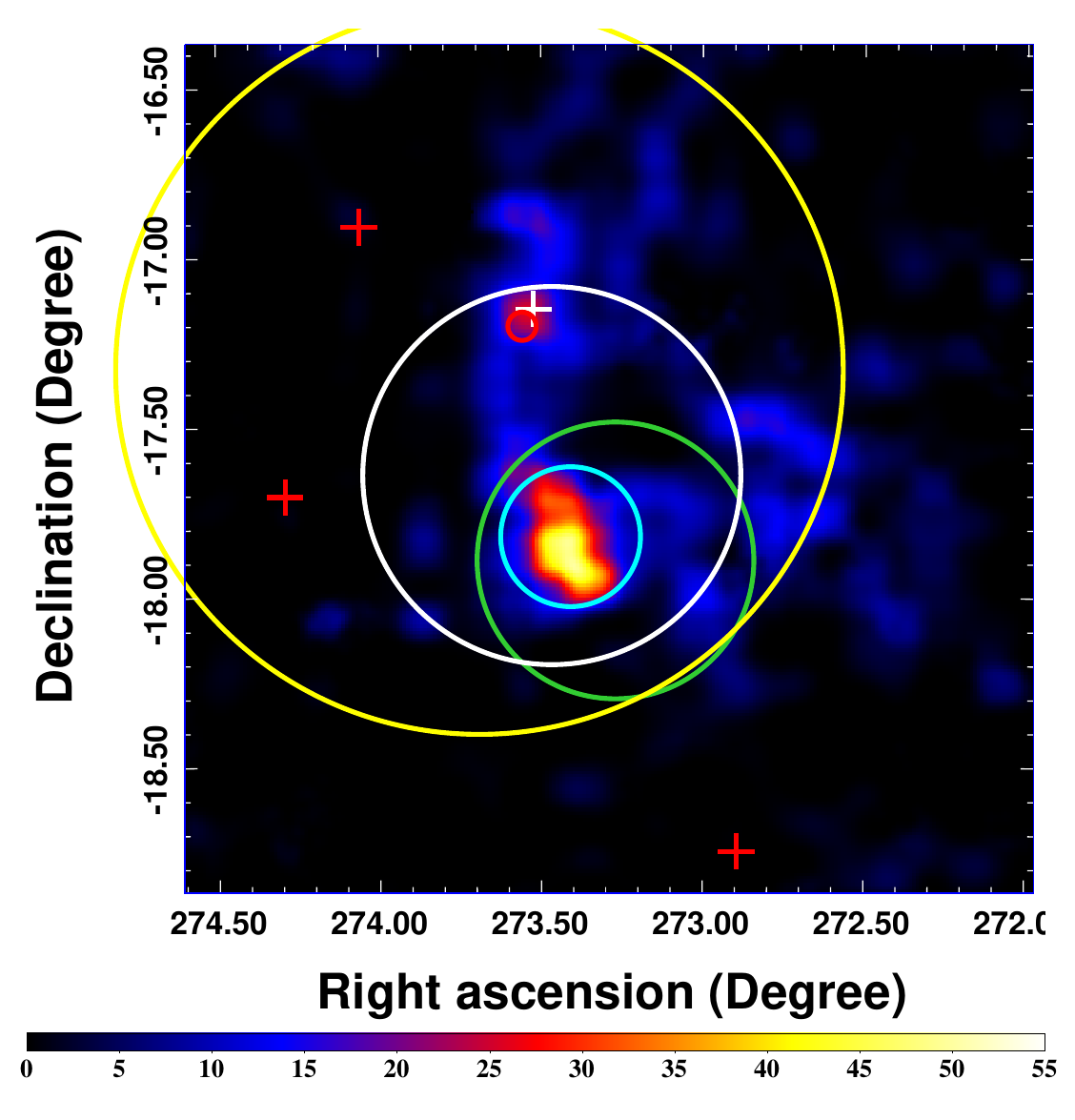}
	\caption{2.5$^\circ \times 2.5^\circ$ TS maps in the energy range of 500 MeV - 10 GeV with ``PSF3'' events (left) and 10 GeV - 1 TeV with ``Source'' events (right). The white and cyan circles represent the spatial size (R$_{\rm 68}$) of SrcA and SrcB in the two energy bands. The positions of 4FGL-DR4 sources are shown as the red crosses. In the left panel, the magenta dashed circle shows R$_{\rm 68}$ of 4FGL J1813.1-1737e in 4FGL-DR4 catalog. In the right panel, the white cross shows the best-fit position of SrcC, and the associated SNR G13.5+0.2 is marked by the red circle.  The yellow circle represents R$_{\rm 68}$ of 1LHAASO J1814-1719u observed by WCDA of LHAASO, and the upper limit of extension given by KM2A above 25 TeV is shown as the green circle\citep{2023arXiv230517030C}.}
\label{fig2:tsmap}
\end{figure*}

To test whether the upturn spectrum is intrinsic or is due to two overlapping components, we do the same likelihood fitting using the events with energies of 500 MeV - 10 GeV (low energy) and 10 GeV - 1 TeV (high energy), respectively.
In the low energy range, we only select the ``PSF3'' events with the best quality of the reconstructed direction to reduce the uncertainties caused by large PSF.
In the high energy range, all events are collected in order to have sufficient counts for the detailed analysis.
For each analysis, we first create a TS map with all sources (except 4FGL J1813.1-1737e) included in the best-fit model (hence
subtracted from the map), which are shown in Figure \ref{fig2:tsmap}.
The TS maps in the low and high energy bands show a clear difference of the centroids and sizes of the $\gamma$-ray emission. 
We thus expect two different $\gamma$-ray components in this region, which are labeled as SrcA for the low energy band and SrcB for the high energy band.
To determine the morphology of the $\gamma$-ray emission from SrcA/SrcB, we used {\tt Fermipy} to test the size of the uniform disk with the events of 500 MeV - 10 GeV (``PSF3'') and 10 GeV - 1 TeV (``Source''), respectively.
The spectrum of SrcA or SrcB are first assumed to be a power-law (PL; dN/dE $\propto$ E$^{-\alpha}$) model.
In the low energy band, the centroid position and R$_{\rm 68}$ of SrcA are fitted to be R.A. = $273^{\circ}\!.467 \pm 0^{\circ}\!.019$, decl. = $-17^{\circ}\!.640 \pm 0^{\circ}\!.018$, and $0^{\circ}\!.557 ^{+0^{\circ}\!.017}_{-0^{\circ}\!.017}$, respectively. The $\gamma$-ray spectrum of SrcA in the low energy band is soft with an index of 2.55 $\pm$ 0.04.
While in the high energy band, the best-fit centroid position and R$_{\rm 68}$ of SrcB are R.A. = $273^{\circ}\!.409 \pm 0^{\circ}\!.026$, decl. = $-17^{\circ}\!.820 \pm 0^{\circ}\!.023$, and $0^{\circ}\!.206 ^{+0^{\circ}\!.017}_{-0^{\circ}\!.017}$, respectively. 
The spatial extension of SrcB is more compact than that of SrcA, and the GeV $\gamma$-ray emission of SrcB concentrates in the TeV $\gamma$-ray emission region of HESS J1813-178 \citep{2006ApJ...636..777A}, while the fitted $\gamma$-ray spectrum of SrcB in the high energy band is hard with an index of 1.97 $\pm$ 0.17.
In addition, the TS map in the high energy range shows an additional point source in the north region of SrcB, which is not included in 4FGL-DR4 catalog, but is in 4FGL-DR2 catalog \citep[4FGL J1814.1-1710;][]{2020arXiv200511208B}. The point source (hereafter labeled as SrcC) is suggested to be associated with SNR G13.5+0.2 in 4FGL-DR2 catalog.
Then we use the updated model including the two extended spatial templates of SrcA and SrcB (Model 2) and the additional SrcC (Model 3) to fit the data in the energy range of 500 MeV - 1 TeV. 
Meanwhile, the models for 4FGL J1813.1-1737e and SrcA/B with two dimensional Gaussian template (Model 4-6) are also tested, and the overall maximum likelihood values of the different models are listed in Table \ref{table:spatial}.
We adopt the Akaike information criterion \citep[AIC;][]{1974ITAC...19..716A}\footnote{AIC = -2ln$\mathcal{L}$+2{\em k}, where $\mathcal{L}$ is the value of maximum likelihood and {\em k} is the parameter numbers of model.} to compare the different models by calculating $\Delta$AIC, the difference between Model 1 and Model 2-6.
It is evident from Table \ref{table:spatial} that the model including the two uniform disks for SrcA and SrcB and the additional point source SrcC provides the minimum $\Delta$AIC value, which is adopted as the spatial model in the following spectral analysis.

\begin{figure*}[!htb]
	\centering
	\includegraphics[width=0.5\textwidth]{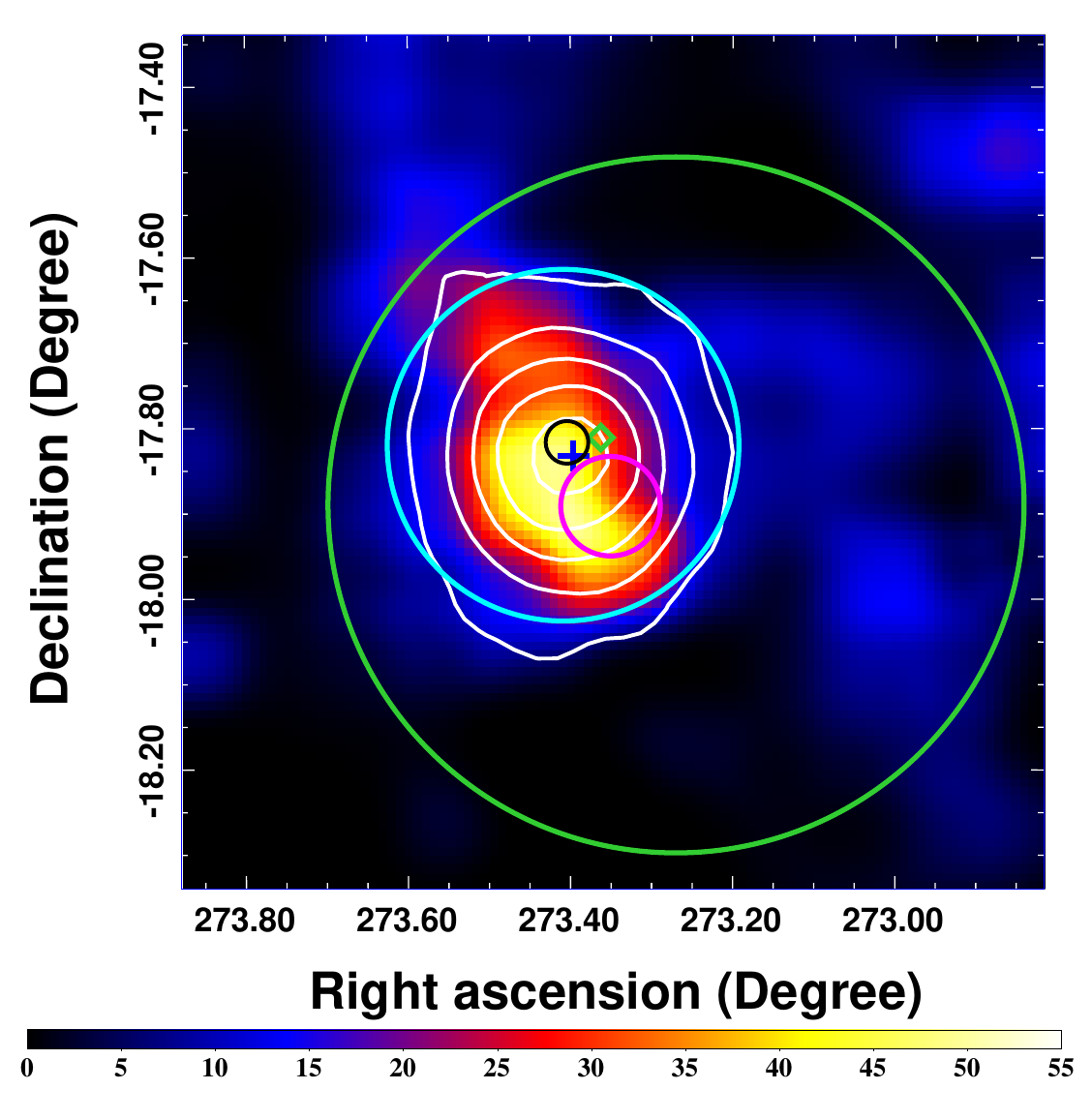}
	\caption{Zoom-in of the TS map for a region of 1.0$^\circ \times 1.0^\circ$ with the energy of 10 GeV - 1 TeV. The cyan circle shows R$_{\rm 68}$ of SrcB. The upper limit size of 1LHAASO J1814-1719 observed by KM2A is marked by the green circle. The black solid and dashed circles show the radio sizes of SNR G12.82-0.02 and SNR G12.72-0.00, respectively. The region of the stellar cluster Cl 1813-178 is marked by the magenta circle \citep{2011ApJ...733...41M}. The position of PSR J1813-1749 is shown as the blue cross \citep{2009ApJ...700L.158G}. The centroid position of the TeV $\gamma$-ray emission from HESS J1813-178 detected by MAGIC \citep{2006ApJ...637L..41A} is marked by the green diamond, and the white contours represent the TeV $\gamma$-ray emission detected by H.E.S.S. \citep{2018A&A...612A...1H}.}
\label{fig3:zoomin-tsmap}
\end{figure*}

\begin{table}[!htb]
\centering
\caption {Spatial Analysis Results between 500 MeV and 1 TeV}
\resizebox{1.0\textwidth}{!}{
\begin{tabular}{ccccccc}
\hline \hline
Spatial Template    &  R.A., Decl.  & R$_{\rm 68}$   & Degrees of Freedom   & -log(Likelihood)  & $\Delta$AIC \\
\hline
%Single disk (4FGL)  & \makecell[c]{$273^{\circ}\!.290$, $-17^{\circ}\!.620$}  & \makecell[c]{$0^{\circ}\!.492$}  & 5  & -6277390.4  & -- \\
%\hline
Single uniform disk (Model 1) & \makecell[c]{$273^{\circ}\!.419 \pm 0^{\circ}\!.014$, $-17^{\circ}\!.625 \pm 0^{\circ}\!.015$ }  & \makecell[c]{$0^{\circ}\!.475 ^{+0^{\circ}\!.015}_{-0^{\circ}\!.014}$}  & 5  & -6277434.8  & 0 \\
\hline
Two uniform disks (Model 2)   & \makecell[c]{SrcA: $273^{\circ}\!.467 \pm 0^{\circ}\!.019$, $-17^{\circ}\!.640 \pm 0^{\circ}\!.018$ \\SrcB: $273^{\circ}\!.409 \pm 0^{\circ}\!.026$, $-17^{\circ}\!.820 \pm 0^{\circ}\!.023$}  & \makecell[c]{SrcA: $0^{\circ}\!.557 ^{+0^{\circ}\!.017}_{-0^{\circ}\!.017}$ \\SrcB: $0^{\circ}\!.206 ^{+0^{\circ}\!.017}_{-0^{\circ}\!.017}$}  & 10  & -6277459.5  & -39.4 \\
\hline
Two uniform disks + Point Source (Model 3)  & \makecell[c]{SrcC: $273^{\circ}\!.523 \pm 0^{\circ}\!.021$, $-17^{\circ}\!.149 \pm 0^{\circ}\!.021$}  & \makecell[c]{--}  & 14  & -6277473.7  & -59.8 \\
\hline
Single Gaussian Template (Model 4) & \makecell[c]{$273^{\circ}\!.454 \pm 0^{\circ}\!.014$, $-17^{\circ}\!.634 \pm 0^{\circ}\!.014$ }  & \makecell[c]{$0^{\circ}\!.517 ^{+0^{\circ}\!.017}_{-0^{\circ}\!.016}$}  & 5  & -6277450.0  & -30.4 \\
\hline
Two Gaussian Templates (Model 5)  & \makecell[c]{SrcA: $273^{\circ}\!.505 \pm 0^{\circ}\!.120$, $-17^{\circ}\!.625 \pm 0^{\circ}\!.120$ \\SrcB: $273^{\circ}\!.331 \pm 0^{\circ}\!.108$, $-17^{\circ}\!.822 \pm 0^{\circ}\!.082$}  & \makecell[c]{SrcA: $0^{\circ}\!.606 ^{+0^{\circ}\!.029}_{-0^{\circ}\!.025}$ \\SrcB: $0^{\circ}\!.235 ^{+0^{\circ}\!.029}_{-0^{\circ}\!.005}$}  & 10  & -6277456.0  & -42.4 \\
\hline
Two Gaussian Templates + Point Source (Model 6) & \makecell[c]{SrcC: $273^{\circ}\!.538 \pm 0^{\circ}\!.019$, $-17^{\circ}\!.150 \pm 0^{\circ}\!.028$}  & \makecell[c]{--}  & 14  & -6277467.5  & -47.4 \\
\hline
\end{tabular}}
\tablecomments{In Model 3/6, the centered positions and R$_{\rm 68}$ of SrcA and Src B are same to that in Model 2/5.}
\label{table:spatial}
\end{table}

\subsection{Spectral Analysis}
\label{spectral}

With the best-fit spatial model including SrcA/B/C in this region, we compared fits using power-law and log-parabola (LogPb; dN/dE $\propto$ E$^{\rm -(\alpha+\beta log(E/E_{\rm b}))}$) spectral functions to find the best model among the two for SrcA/B in the energy range of 500 MeV - 1 TeV.
SrcC is spatially coincident with SNR G13.5+0.2, not with HESS J1813-178, and its spectrum is fixed to be a power-law model.
The spectral index of SrcC is fitted to be $2.65 \pm 0.29$, with the integrated photon flux of $(4.43 \pm 1.21)\times10^{-9}$ photon cm$^{-2}$ s$^{-1}$.
The fitting results with the different spectral models for SrcA/B are listed in Table \ref{table:spectra}. 
We compared the overall likelihood values with the PL models for SrcA/B, and the variation suggests an evidence of spectral curvature for the $\gamma$-ray spectrum of SrcA with TS$_{\rm curve}$=2($\ln\mathcal{L}_{\rm LogPb-PL}$-$\ln\mathcal{L}_{\rm PL-PL}$) = 9.0, which corresponds to a significance of $\sim$3.0$\sigma$ with one additional degree of freedom.
The fitting of the LogPb model for SrcA gives the spectral parameters of $\alpha = 2.52 \pm 0.10$ and $\beta = 0.10 \pm 0.05$. The integrated photon flux of SrcA is calculated to be $(3.71 \pm 0.35)\times10^{-8}$ photon cm$^{-2}$ s$^{-1}$ with the TS value of 1358.9.
For SrcB, a power-law spectrum is enough to describe its $\gamma$-ray emission with the spectral index of $2.11 \pm 0.08$. The corresponding integrated photon flux is calculated to be $(8.72 \pm 1.48)\times10^{-9}$ photon cm$^{-2}$ s$^{-1}$ with the TS value of 270.6.

\begin{table}[!htb]
\centering
\caption{Comparison of the different Spectral Models (500 MeV - 1 TeV)}
\begin{tabular}{cccccccc}
\hline\hline
SrcA & SrcB  & Degrees of Freedom & -log(Likelihood) &TS$_{\rm curve}$\\
\hline
PL    & PL      & 4  & -6277473.7 &  0.0   \\
LogPb & PL      & 5  & -6277478.2 &  9.0   \\
PL    & LogPb   & 5  & -6277473.7 &  0.0   \\
LogPb & LogPb   & 6  & -6277476.4 &  5.4   \\
\hline
\end{tabular}
\tablecomments{TS$_{\rm curve}$ is the difference between the TS values obtained for each spectral function.}
\label{table:spectra}
\end{table}

\begin{figure*}[!htb]
	\centering
    \includegraphics[width=0.32\textwidth]{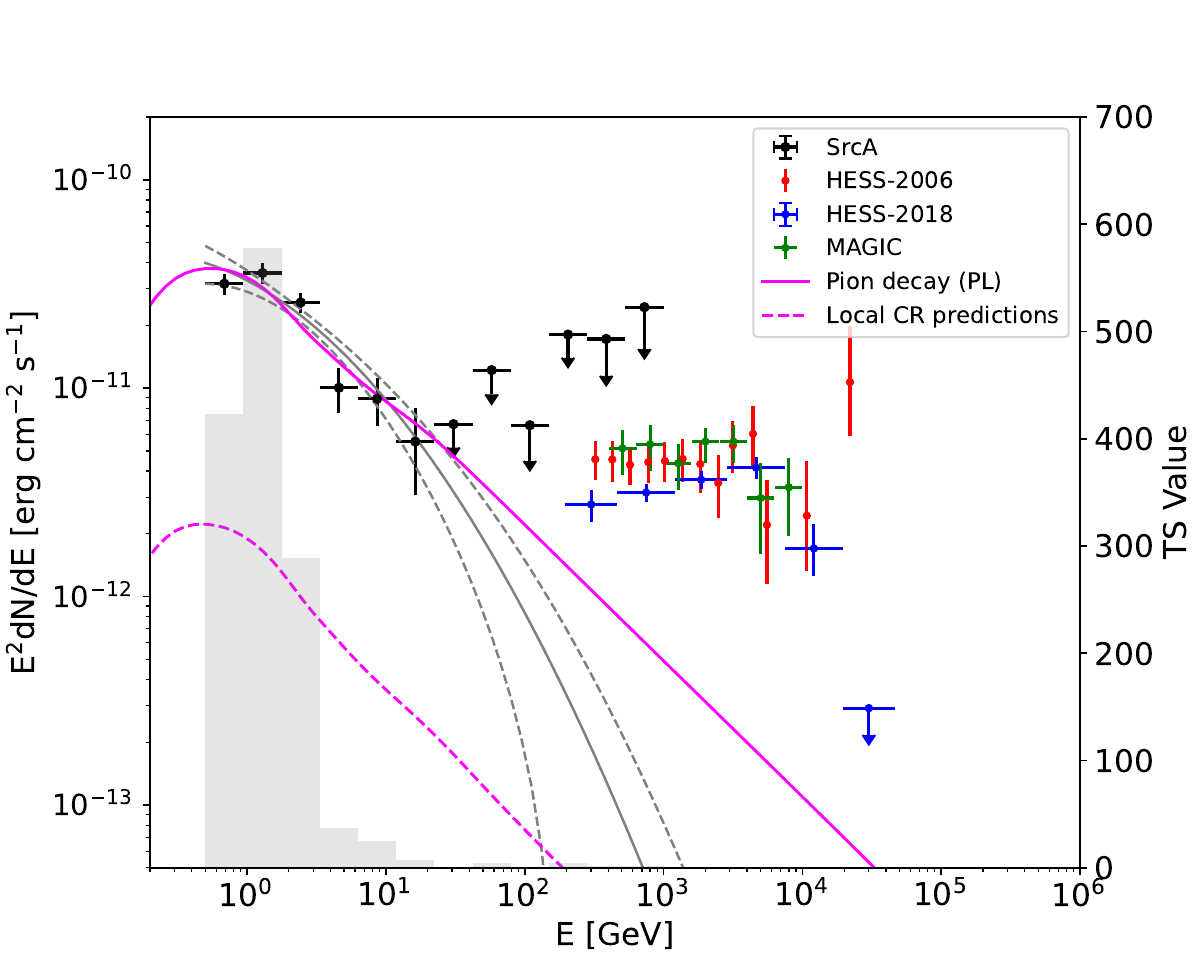}
     \includegraphics[width=0.32\textwidth]{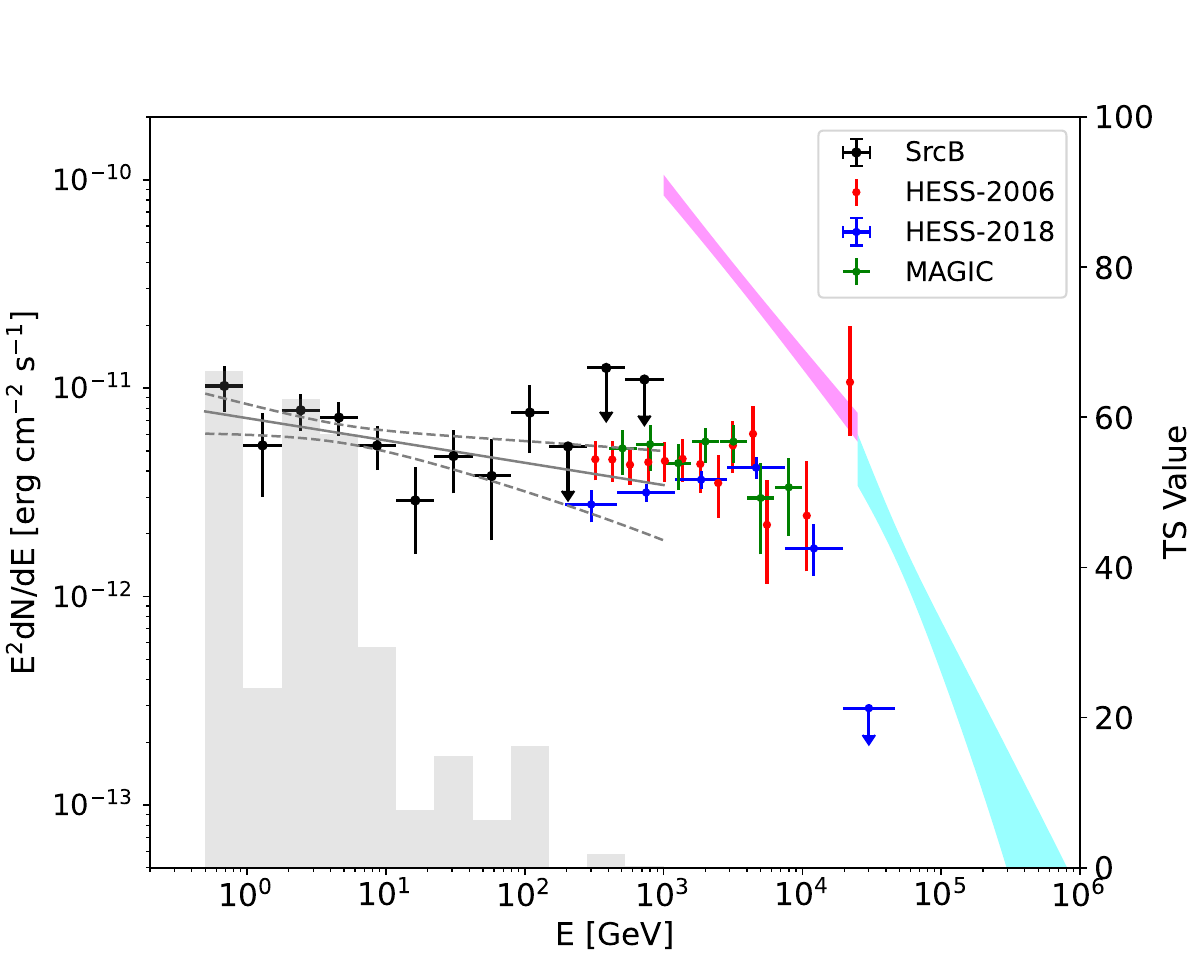}
      \includegraphics[width=0.32\textwidth]{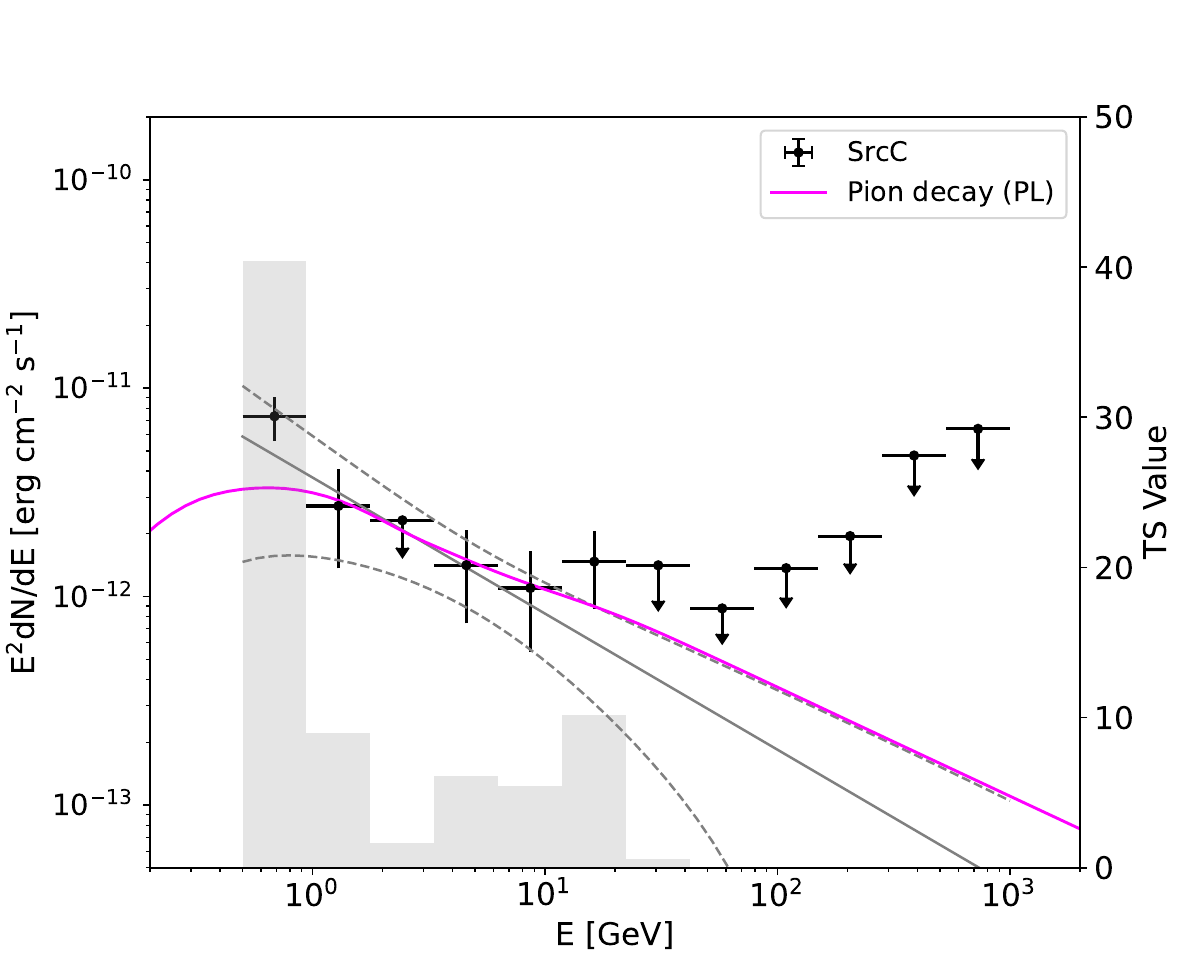}
\caption{The SEDs of SrcA (left), SrcB (middle) and SrcC (right). The black dots depict the results of {\em Fermi}-LAT data in the energy range of 500 MeV - 1 TeV, together with the global best-fit spectra with 1$\sigma$ propagated statistical errors shown as dashed gray lines. The black arrows indicate the 95\% upper limits and the gray histogram shows the TS value for each energy bin. The TeV $\gamma$-ray data of HESS J1813-178 observed by H.E.S.S. \citep{2006ApJ...636..777A, 2018A&A...612A...1H} and MAGIC \citep{2006ApJ...637L..41A} are marked by the red, blue and green dots, as shown in the legend. In the middle panel, the magenta and cyan filled butterflies present the spectra of 1LHAASO J1814-1719u observed by WCDA and KM2A of LHAASO, respectively \citep{2023arXiv230517030C}. In the left and right panels, the magenta solid lines show the hadronic models by adopting a power-law spectrum of protons for SrcA and SrcC, respectively. And 
the dashed magenta line in the left panel represents the predicted $\gamma$-ray emission assuming that the CR spectra therein is the same as that measured locally by AMS-02 \citep{2015PhRvL.114q1103A}.}
\label{fig4:three sed}
\end{figure*}

To obtain the SED of SrcA/B/C, we divide the data from 500 MeV to 1 TeV into 12 bins in energy equally-spaced logarithmically, and perform the same likelihood fitting analysis for each energy bin.
For the energy bin with TS value of SrcA/B/C lower than 5.0, an upper limit with 95\% confidence level is calculated.
The results of the SED are shown in Figure \ref{fig4:three sed}, and the global fitting with the PL or LogPb models are also overplotted.
The SED of SrcA shows a soft GeV $\gamma$-ray spectrum, which is not consistent with the TeV $\gamma$-ray SED of HESS J1813-178.
While the hard GeV $\gamma$-ray spectrum of SrcB smoothly connects with the TeV spectrum of HESS J1813-178.
Also considering the spatial coincidence and compact $\gamma$-ray emission region between SrcB and HESS J1813-178 shown in Figure \ref{fig3:zoomin-tsmap}, we suggest that SrcB is the GeV counterpart of HESS J1813-178.

\section{CO observations}
\label{CO-observation}

To search for molecular clouds, spatially and morphologically coincident with the $\gamma$-ray emission around HESS J1813-178, we analyzed the $^{\rm 12}$CO($J$=1-0) and $^{\rm 13}$CO($J$=1-0) line data observed by the FOREST Unbiased Galactic plane Imaging survey with the Nobeyama 45 m telescope \citep[FUGIN;][]{2017PASJ...69...78U}.
The Nobeyama telescope has a beam size of 14" at 115 GHz, which results in a postreduction $^{\rm 12}$CO($J$=1-0) angular resolution of 20" and a $^{\rm 13}$CO($J$=1-0) angular resolution of 21" for FUGIN survey data with the velocity resolution of 1.3 km s$^{\rm -1}$.

\begin{figure*}[!htb]
\centering
	\includegraphics[width=0.32\textwidth]{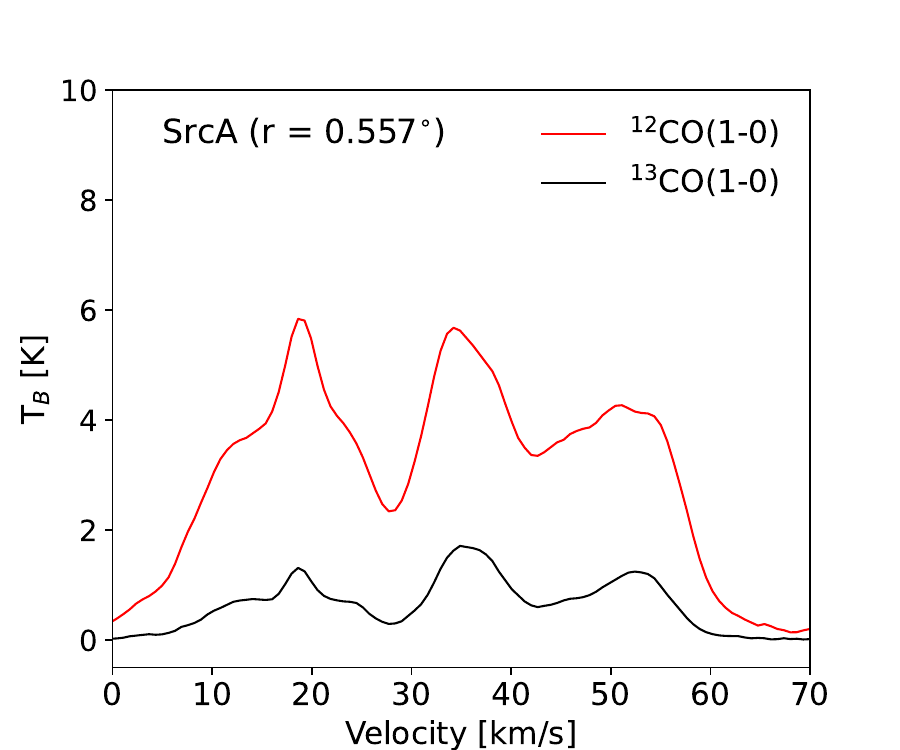}
        \includegraphics[width=0.32\textwidth]{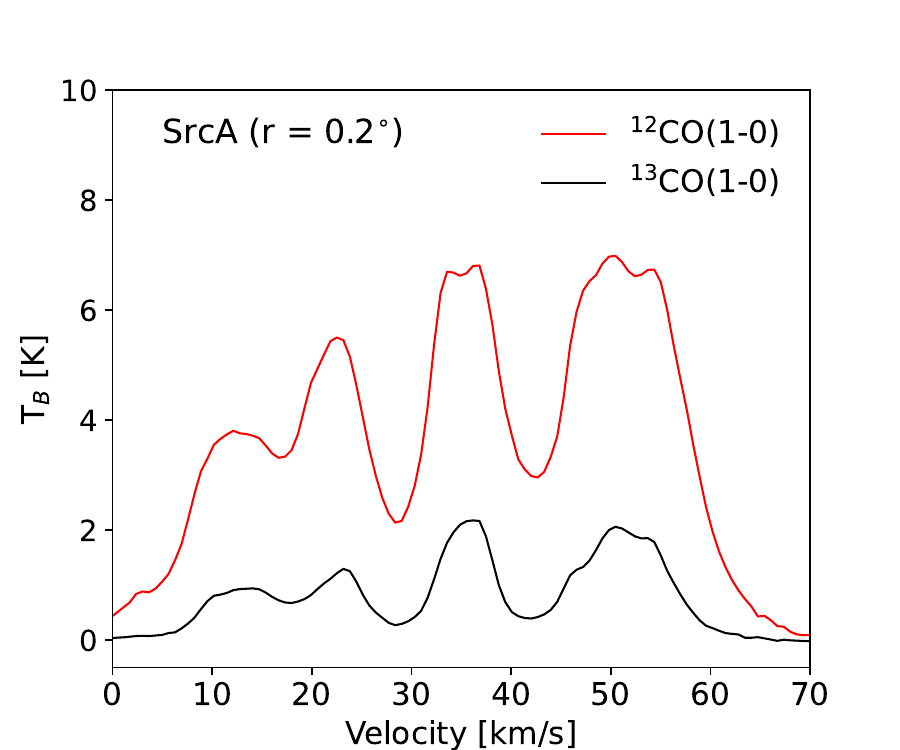}
        \includegraphics[width=0.32\textwidth]{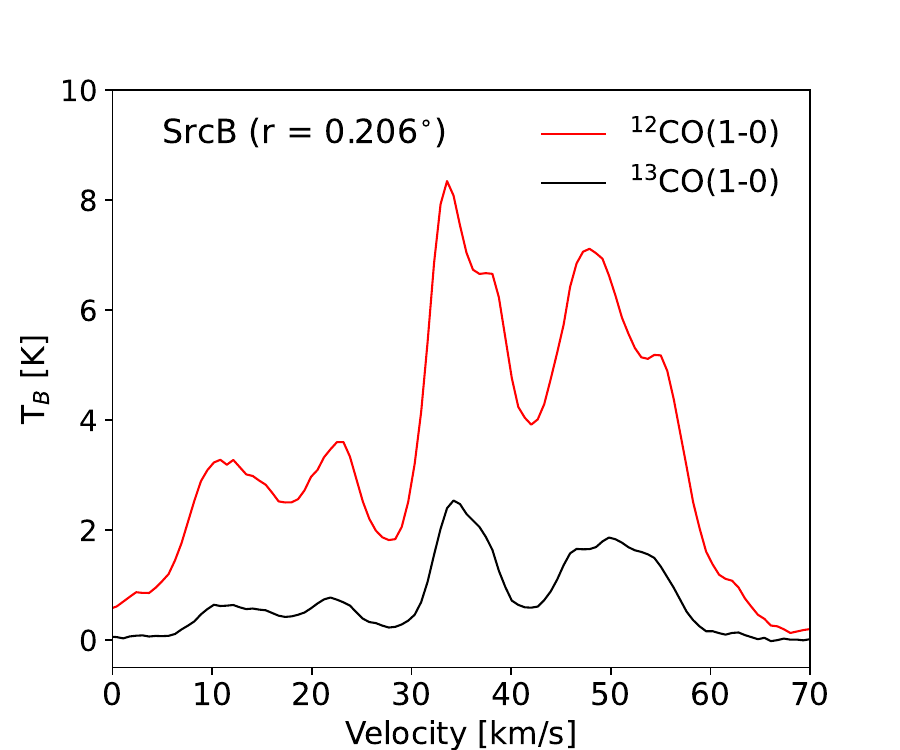}\\
        \includegraphics[width=0.32\textwidth]{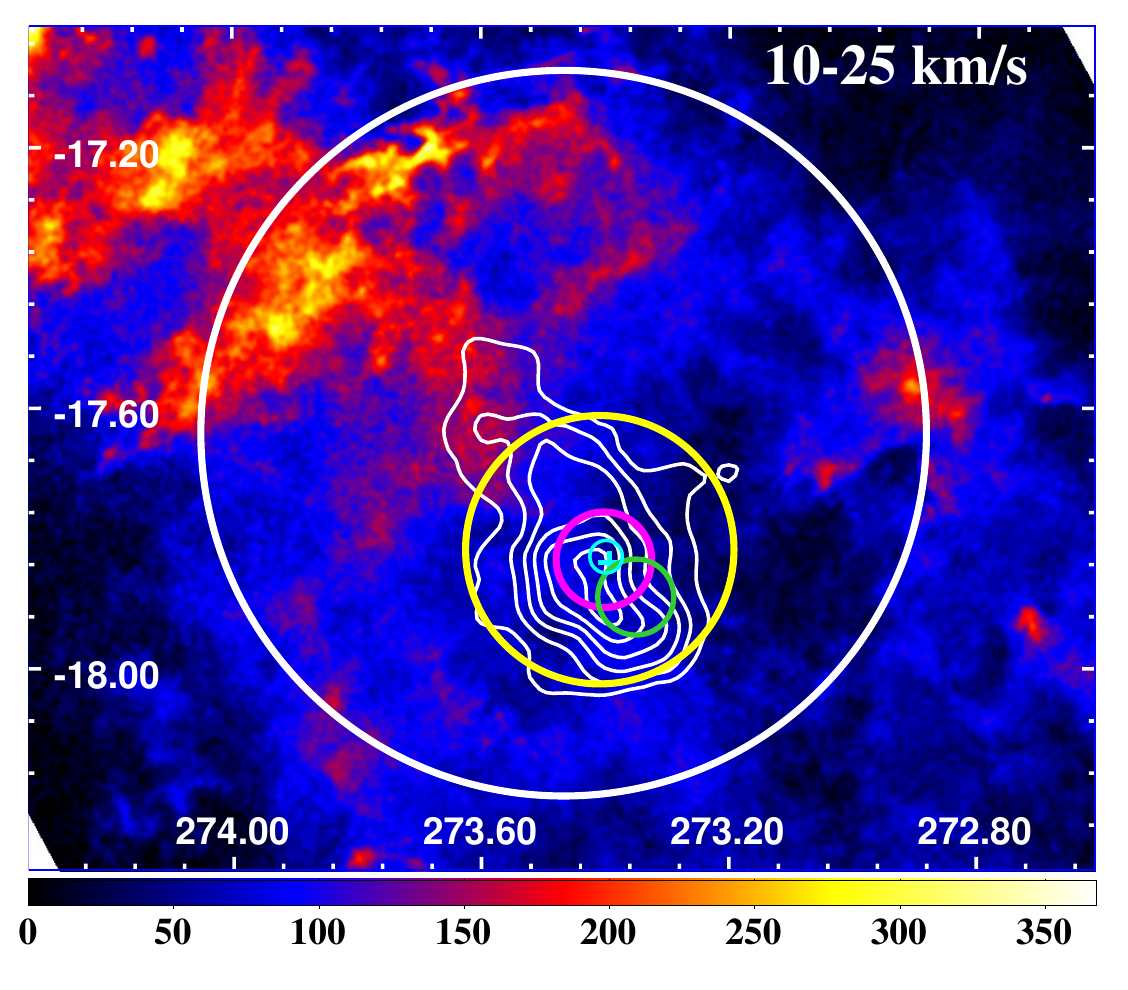}
        \includegraphics[width=0.32\textwidth]{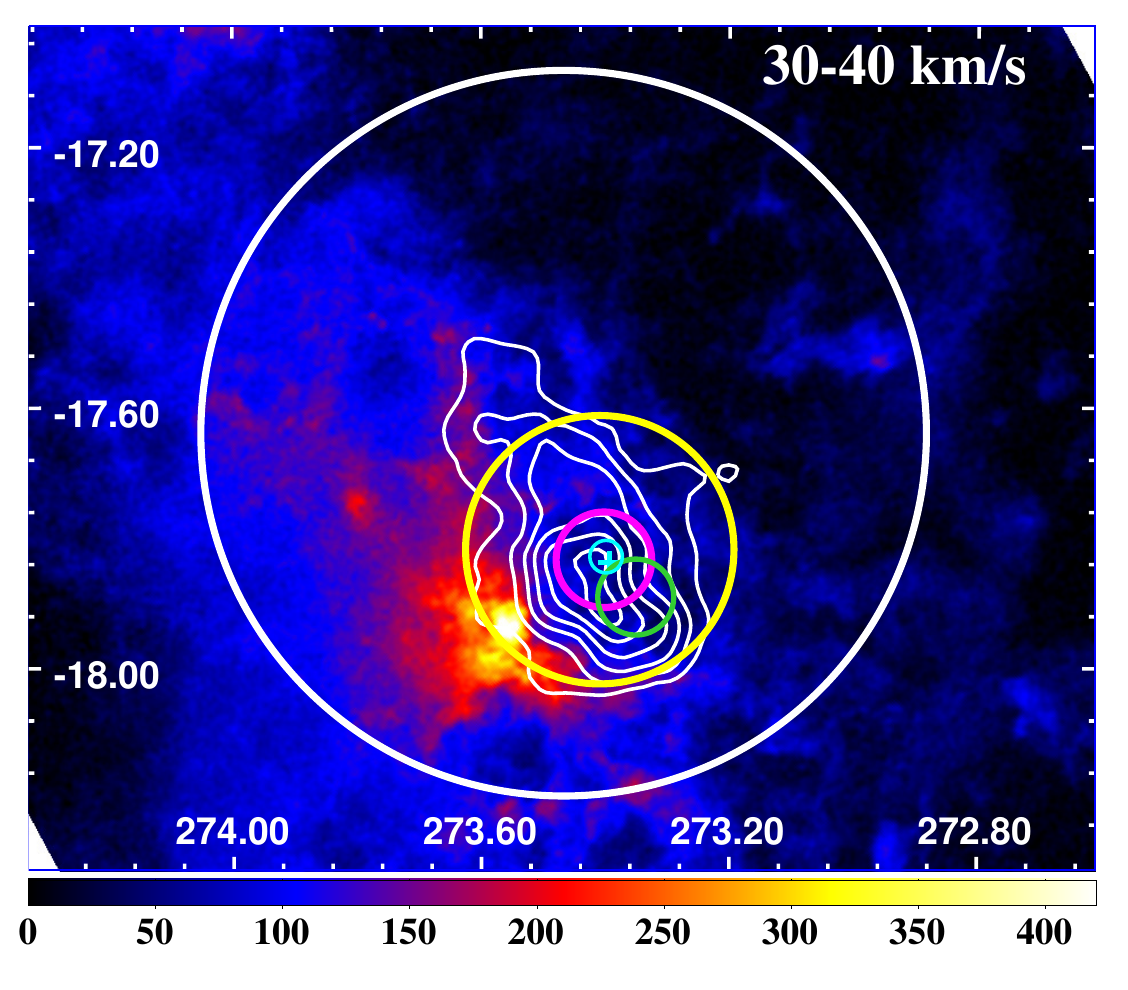}
        \includegraphics[width=0.32\textwidth]{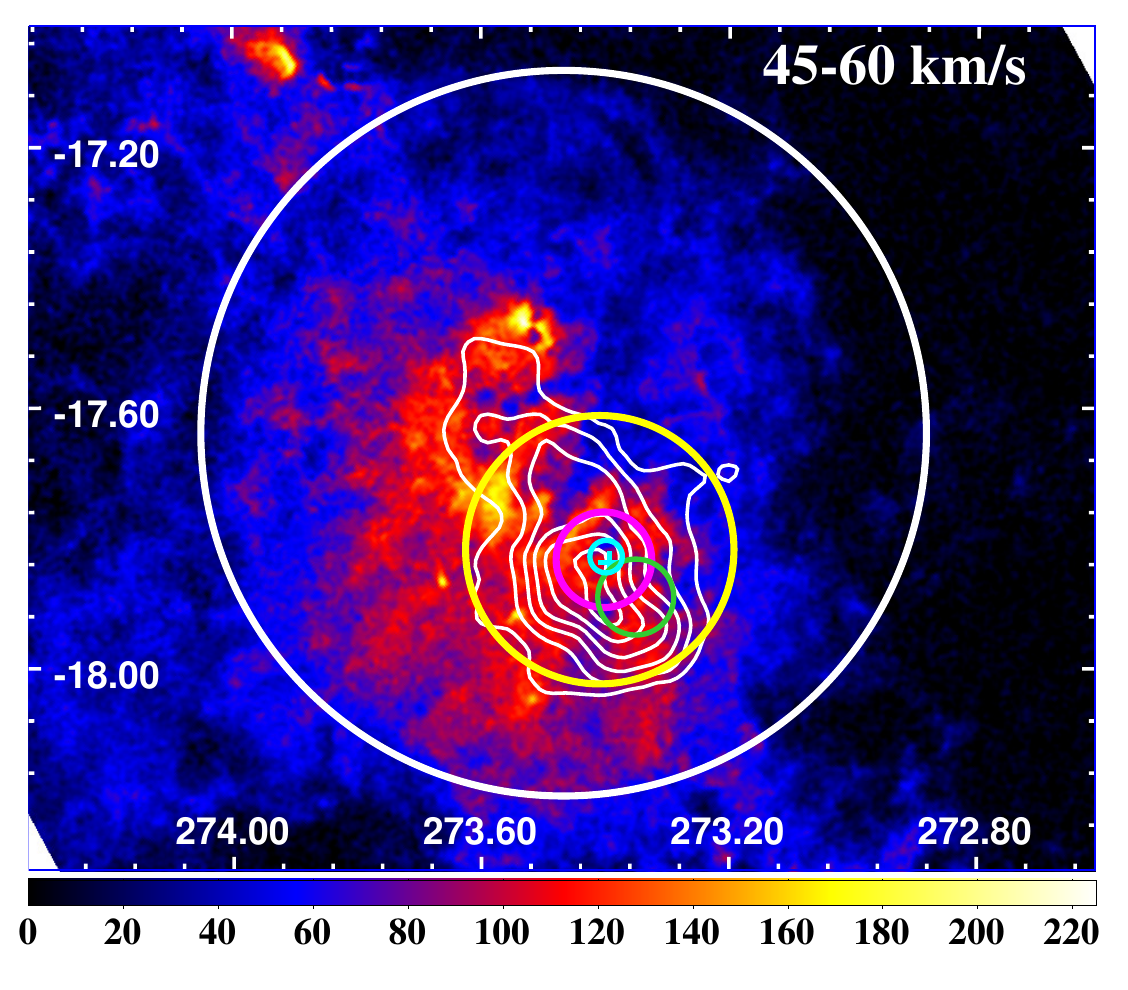}
\caption{Top: $^{\rm 12}$CO($J$=1-0; red) and $^{\rm 13}$CO($J$=1-0; black) spectra of molecular gas toward SrcA within a region of r = $0^{\circ}\!.557$ (left), r = $0^{\circ}\!.2$ (middle), and HESS J1813-178 within a region of r = $0^{\circ}\!.206$ (right). Bottom: $^{12}$CO($J$ = 1–0) intensity maps in the velocity ranges of 10-25 km s$^{-1}$ (left), 30-40 km s$^{-1}$ (middle), and 45-60 km s$^{-1}$ (right). The white and yellow circles show R$_{\rm 68}$ of SrcA and SrcB, respectively. The radio size of SNR G12.82-0.02 and the region of stellar cluster Cl 1813-178 are marked by the cyan and green circles, together with the position of PSR J1813-1749 shown as the cyan cross. The magenta circle shows the TeV $\gamma$-ray extension of HESS J1813-178 \citep{2018A&A...612A...1H}. The white contours correspond to the GeV $\gamma$-ray emission of HESS J1813-178 in the energy range of 10 GeV - 1 TeV by {\em Fermi}-LAT.}
\label{CO-Velocity2Temp}
\end{figure*}

The spectra of $^{\rm 12}$CO($J$=1-0) and $^{\rm 13}$CO($J$=1-0) emission are extracted toward SrcA within a radius of $0^{\circ}\!.557$ and $0^{\circ}\!.2$, and HESS J1813-178 within a radius of $0^{\circ}\!.206$, which are shown in the top panels of Figure \ref{CO-Velocity2Temp}.
The spectra display three distinctive peaks in the velocity intervals of $\sim$ 10-25, 30-40, and 45-60 km s$^{-1}$, and the $^{12}$CO($J$ = 1–0) intensity maps in the corresponding ranges are shown in the bottom panels of Figure \ref{CO-Velocity2Temp}.
Note that the molecular cloud at 30-40 km s$^{-1}$ has been studied by \citet{2007A&A...470..249F} with the NANTEN $^{\rm 12}$CO($J$=1-0) observations, which is suggested to be related with the star-forming region W33.
%It can be seen that the molecular gas in the velocity ranges of 10-25 and 30-40 km s$^{-1}$ are located far from the peak of the $\gamma$-ray emission from SrcA or HESS J1813-178, which makes the association unlikely.
Although the molecular gas in the velocity ranges of 10-25 and 30-40 km s$^{-1}$ are partly overlapping with the $\gamma$-ray emission, the wide distribution of molecular gas in the range of 45-60 km/s in the $\gamma$-ray emission region makes the correlation more likely for the scenario in which the $\gamma$-rays are produced by interactions of cosmic rays with the gas.
%The intensity of the $\gamma$-ray emission from SrcA or HESS J1813-178 spatially correlate more strongly with the gas distribution in the velocity range of 45-60 km s$^{-1}$.

Adopting the standard Galactic rotation model in \citet{2014ApJ...783..130R}, we calculated the possible kinematic distances associated with the velocity range of 45-60 km s$^{-1}$.
It should be noted that the velocity of each MC could indicate two candidate kinematic distances, the near-side one and the far-side one.
The near and far distances of the MC in 45-60 km s$^{-1}$ are calculated to be 3.9-4.6 kpc and 11.5-12.1 kpc, 
which are in accord with the values of 4.8 $\pm$ 0.3 kpc for the young stellar cluster, Cl 1813-178, and the higher value of 12 kpc for PSR J1813-1749/SNR G12.82-0.02. 

%Figure \ref{COmap} displays the $^{\rm 12}$CO($J$=1-0) and $^{\rm 13}$CO($J$=1-0) maps for ten consecutive velocity ranges from 26 to 66 km s$^{-1}$, with a step of 4 km s$^{-1}$ for each map. The molecular gas are observed to be spatially associated with the GeV $\gamma$-ray emission in the velocity range from 42-62 km s$^{-1}$. The correlation is the best in the velocity ranges of 46-50 and 50-58 km s$^{-1}$ with the $\gamma$-ray emission from HESS J1813-178 and SrcA, respectively.

Adopting the CO-to-H$_{\rm 2}$ conversion factor $X_\mathrm{CO} = 2\times10^{20}$ cm$^{-2}$ (K km s$^{-1})^{-1}$ \citep{2013ARA&A..51..207B}, we calculated the total mass of molecular gas within the 68\% containment radius of extended GeV $\gamma$-ray emission for SrcA (r = $0^{\circ}\!.557$) and HESS J1813-178 (r = $0^{\circ}\!.206$) with the different distances.
For the distance of 4.8 kpc, the total masses of gas within SrcA and HESS J1813-178 are estimated to be $\sim$ $1.82\times 10^{6}d_{4.8}^{2} M_{\odot}$ and $\sim$ $3.86\times 10^{5}d_{4.8}^{2} M_{\odot}$, respectively. The corresponding average gas number densities are about $\rm{n_{\rm gas}}$ = 170$d_{4.8}^{-1}$ cm$^{-3}$ and 720$d_{4.8}^{-1}$ cm$^{-3}$  by assuming a spherical geometry of the gas distribution.
For the distance of 12 kpc, the calculated total masses of gas within SrcA and HESS J1813-178 are $\sim$ $1.14\times 10^{7}d_{12}^{2} M_{\odot}$ and $\sim$ $2.41\times 10^{6}d_{12}^{2} M_{\odot}$ with the corresponding average gas number densities of 70$d_{12}^{-1}$ cm$^{-3}$ and 300$d_{12}^{-1}$ cm$^{-3}$.

\section{Discussion}
\label{dis}

The spatial and spectral data analyses with the different energy ranges reveal that the diffuse GeV $\gamma$-ray emission around HESS J1813-178 can be distinguished into three separate components. 
SrcA with an extension of $0^{\circ}\!.557$ has a soft GeV $\gamma$-ray spectrum, which has no identified associated counterpart, and a point source in the north of HESS J1813-178, SrcC, is spatially associated with the SNR G13.5+0.2.
For SrcB, the GeV $\gamma$-ray emission region is more compact than SrcA, which is spatially coincident with the TeV $\gamma$-rays of HESS J183-178.
The GeV $\gamma$-ray spectrum of SrcB is hard with an index of 2.11, which also smoothly connects with the TeV spectrum of HESS J1813-178.
Both the spatial morphology of the $\gamma$-ray emission and the GeV-TeV $\gamma$-ray spectrum suggest that SrcB could be the GeV counterpart of HESS J1813-178.
Moreover, the CO observations display that the intensity of the $\gamma$-ray emission from SrcA and HESS J1813-178 spatially correlate with the molecular gas distribution in the velocity range of 45-60 km s$^{\rm -1}$.
Such velocity range indicates two candidate kinematic distances, and the near one is in accord with the distance of the young stellar cluster, Cl 1813-178, of $\sim$ 4.8 kpc.
The far one with $\sim$ 12 kpc is in accord with the distance of PSR J1813-1749/SNR G12.82-0.02 system.

\subsection{The possible $\gamma$-ray origin of HESS J1813-178}
Considering the association between HESS J1813-178 and PSR J1813-1749/SNR G12.82-0.02 or Cl 1813-178 at the different distances, here we discuss these two possibilities as the origin of HESS J1813-178.
The $\gamma$-ray emission can be produced by the decay of neutral pions due to the inelastic nuclei-nuclei collisions, which is the hadronic process.
In addition, the leptonic process, that is the inverse Compton scattering or the bremsstrahlung (Brem) process of relativistic electrons, can also produce the intense $\gamma$-ray emission.
For the model fitting, the TeV observations of HESS J1813-178 from H.E.S.S. \citep{2006ApJ...636..777A,2018A&A...612A...1H} and MAGIC \citep{2006ApJ...637L..41A} are adopted. 
The observations by LHAASO-WCDA show a larger extension and a higher flux of the $\gamma$-ray emission around HESS J1813-178 in the energy range of 1-25 TeV \citep{2023arXiv230517030C}, which are not consistent with the results of H.E.S.S. and MAGIC. 
Such results could be due to the contributions from the unresolved $\gamma$-ray sources in the more extended $\gamma$-ray region of LHAASO-WCDA.
LHAASO-KM2A gives an upper limit of the extension of the $\gamma$-ray emission above 25 TeV, and its spectrum also could connect with that of H.E.S.S. and MAGIC.
Therefore, only LHAASO-KM2A data are used in the following model fitting.

\subsubsection{PWN or SNR scenario at the distance of 12 kpc}

\begin{table}[!htb]
	\centering
	\caption {Parameters of the leptonic model as a PWN for HESS J1813-178}
	\resizebox{1.0\textwidth}{!}{
		\begin{tabular}{ccccccccc}
			\hline \hline
			Model  & $\alpha_{\rm e,1}$ &   $\alpha_{\rm e,2}$ &  E$_{\rm e, break}$(GeV)  & E$_{\rm e, cut}$(TeV)  & W$_{\rm e}$(erg)  &  B($\mu$G)  & $\rm n_{gas}(cm^{-3})$ & $\chi^2/N_{\rm dof}$ \\
			\hline
			ICS-dominated (one-zone)   &  $1.12^{+0.20}_{-0.08}$ &  $2.78^{+0.03}_{-0.03}$  &  $143.30^{+41.33}_{-28.20}$ &  $738.62^{+107.36}_{-127.52}$  &  $(1.18^{+0.37}_{-0.20})(\rm d / 12.0 kpc)^{2} \times 10^{49}$ & $7.34^{+0.41}_{-0.36}$ & 1.0  &  110.80/45=2.46 \\
			\hline
			Brem-dominated (one-zone) &  $1.98^{+0.02}_{-0.06}$ &  $2.33^{+0.04}_{-0.05}$  &  $125.40^{+50.78}_{-22.22}$ &  $433.95^{+87.26}_{-83.97}$  &  $(1.22^{+0.15}_{-0.19})(\rm d / 12.0 kpc)^{2} \times 10^{48}$ & $7.78^{+0.42}_{-0.57}$ & 300.0 &  141.07/45=3.13 \\
			\hline
			ICS-dominated (zone 1)   &  $1.64^{+0.16}_{-0.18}$ &  --  &  -- &  $201.35^{+80.41}_{-53.11}$  &  $(7.85^{+1.82}_{-0.69})(\rm d / 12.0 kpc)^{2} \times 10^{46}$ & $8.50^{+1.27}_{-0.99}$ & 1.0  & {\multirow{2}{*}{85.14/43=1.98}}  \\
			%  	 \cline{2-8}
			ICS-dominated (zone 2)       &  -- &  $2.77^{+0.11}_{-0.11}$  & -- &  $89.15^{+16.94}_{-22.21}$  &  $(2.97^{+0.79}_{-0.68})(\rm d / 12.0 kpc)^{2} \times 10^{50}$ & $1.54^{+0.63}_{-0.42}$ & 1.0  &   \\
			\hline \hline
	\end{tabular}}
	\tablecomments{$\alpha_{\rm e,1}$ and $\alpha_{\rm e,2}$ represent the indices below and above E$_{\rm e, break}$ for BPL spectrum of electrons in one-zone model, or the indices for two PL spectra of electrons in two-zone model. And the total energy of electrons, $W_{\rm e}$, is calculated for $E_{\rm e} > 1$ GeV.}
	\label{table:model-leptonic}
\end{table} 

\begin{figure*}[!htb]
	\centering
	\includegraphics[width=0.32\textwidth]{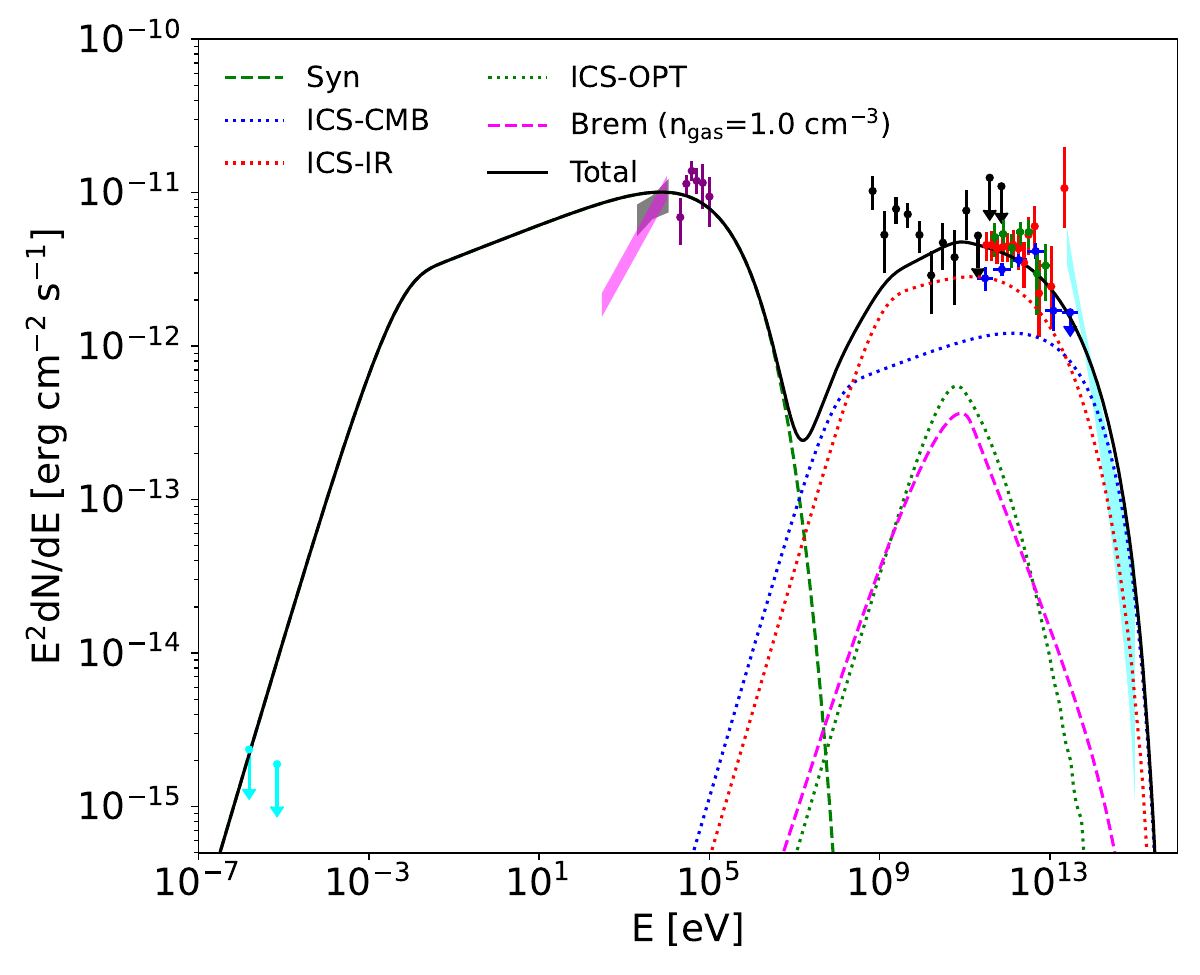}
	\includegraphics[width=0.32\textwidth]{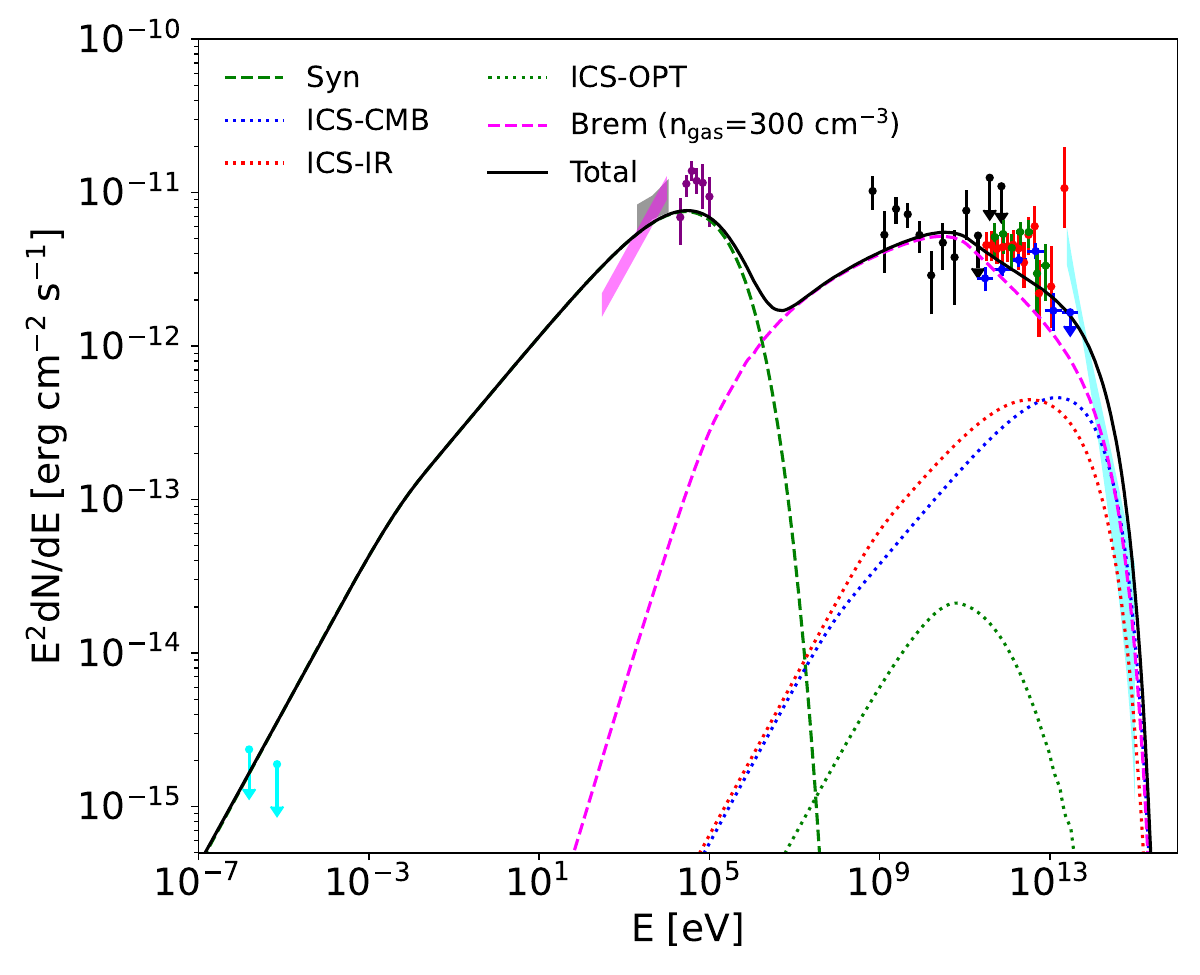}
	\includegraphics[width=0.32\textwidth]{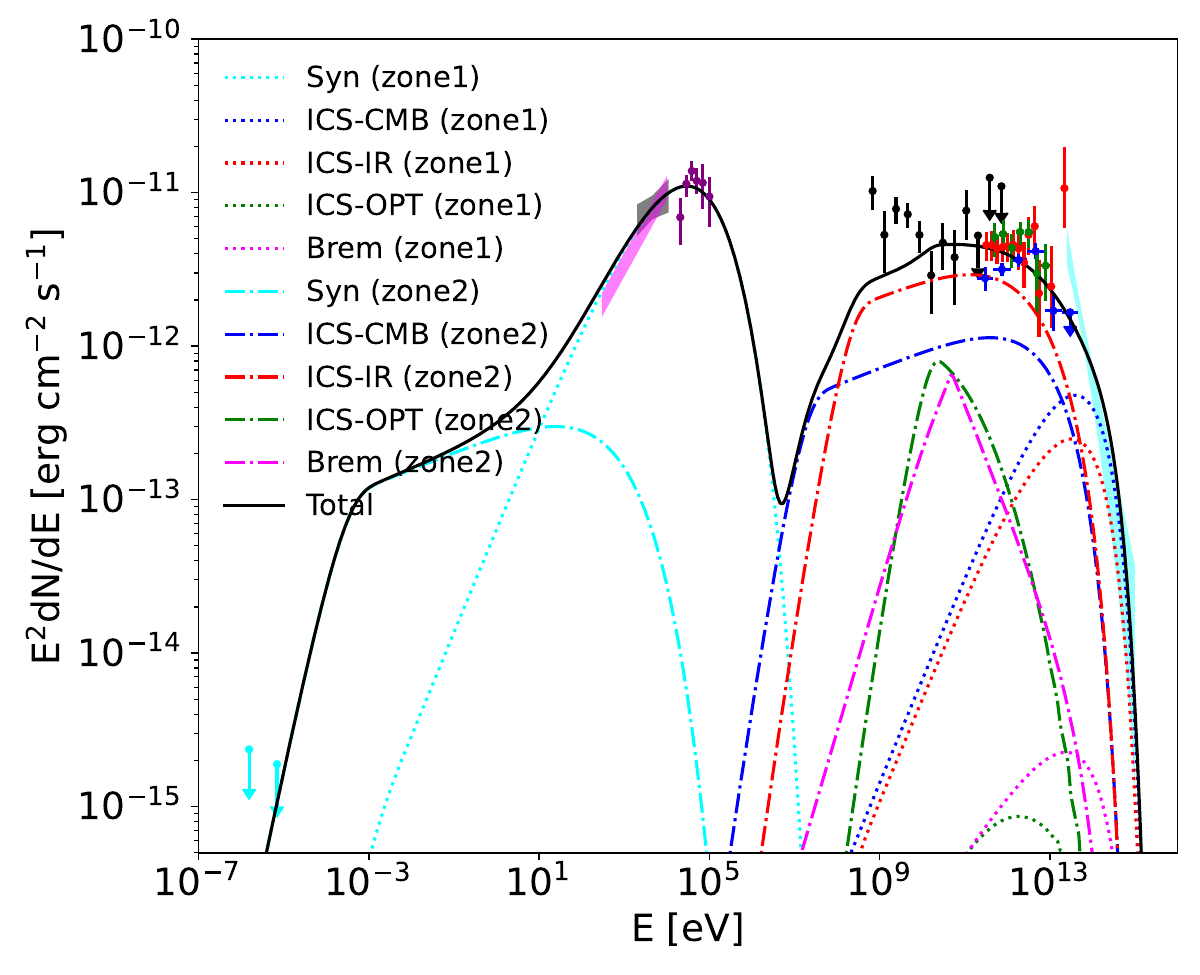}
	\caption{The leptonic model fitting as a PWN for HESS J1813-178 with one-zone ICS-dominated scenario (left), Brem-dominated scenario (middle) and two-zone ICS-dominated scenario (right). The radio upper limits shown as the cyan arrows are from \citet{2007A&A...470..249F}. The gray and magenta butterflies mark the X-ray spectra of PWN observed by \citet{2007A&A...470..249F} and \citet{2020MNRAS.498.4396H}, together with the {\em INTEGRAL} observations shown as the purple dots \citep{2005ApJ...629L.105B}. For the left and middle panels, the green and magenta dashed lines represent the synchrotron and bremsstrahlung components, respectively. The contributions from ICS process with the different radiation fields are shown as the dotted lines, and the black solid line represents the sum of the contributions from different radiation components. For the right panel, the radiation components of zone 1 and zone 2 are marked by the dotted and dot-dashed lines, as shown in the legend. The contributions from ICS-OPT and bremsstrahlung components of zone 1 are too low to show.}
	\label{fig:leptonic}
\end{figure*}

%If HESS J1813-178 is related with PSR J1813-1749, the $>$200 GeV luminosity of HESS J1813-178 would be $\approx 3 \times 10^{35} \rm erg \,s^{-1}$ with this distance, which makes it to be one of the most luminous TeV sources in the Galaxy.
The PWN component associated with PSR J1813-1749 was first resolved by \citet{2007ApJ...665.1297H} using the observations of {\em Chandra}. The spectrum of the PWN follows a power law with an index of $\sim$1.3, and an absorbed flux of 5.6 $\times$ 10$^{\rm -12}$ erg cm$^{\rm -2}$ s$^{\rm -1}$ is obtained in the energy range of 2-10 keV.
Following {\em XMM}-Newton observations also resolved the PWN with an spectral index of $\sim$1.8 and an absorbed flux of 7 $\times$ 10$^{\rm -12}$ erg cm$^{\rm -2}$ s$^{\rm -1}$ from 2 to 10 keV \citep{2007A&A...470..249F}. 
In the high energy band of 20-100 keV, {\em INTEGRAL} discovered a soft $\gamma$-ray source with an index of $\sim$1.8, which is spatially consistent with the PWN component \citep{2005ApJ...629L.105B}.
The newest X-ray data analysis combining {\em Chandra} and {\em NICER} data by \citet{2020MNRAS.498.4396H} modeled the PWN with an spectral index of 1.5 $\pm$ 0.1, which is slight harder than the results of previous works.
And an absorbed 2-10 keV flux of 7.6 $\times$ 10$^{\rm -12}$ erg cm$^{\rm -2}$ s$^{\rm -1}$ is calculated.

We first considered the PWN scenario for the $\gamma$-ray emission of HESS J1813-178.
The spin-down luminosity of PSR J1813-1749 is $\dot{E} = 5.6 \times 10^{37} \ \rm erg \ s^{-1}$ with its characteristic age of $\tau_c = 5600 \rm \ yr$  \citep{2012ApJ...753L..14H, 2020MNRAS.498.4396H}, which is similar to that of PSR J1513-5908 associated with the $\gamma$-ray PWN MSH 15-52 \citep[$\dot{E} = 1.8 \times 10^{37} \ \rm erg \ s^{-1}$ and $\tau_c = 1700 \rm \ yr$;][]{1982ApJ...256L..45S, 2010ApJ...714..927A}. The GeV $\gamma$-ray luminosity of HESS J1813-178 in the energy range of 1 GeV - 1 TeV is calculated to be $L_{\rm 1 GeV-1 TeV} \approx 6.0\times 10^{35}~(d/{\rm 12.0~kpc})^2~\rm{erg}~\rm{s}^{-1}$, and the radius of GeV extension is about 43 pc by adopting the distance of 12 kpc. Such values are about two times higher than that of MSH 15-52, whose GeV $\gamma$-ray luminosity and the radius of GeV extension are calculated to be $L_{\rm 1 GeV-1 TeV} \approx 2.8\times 10^{35}~(d/{\rm 12.0~kpc})^2~\rm{erg}~\rm{s}^{-1}$ and 19 pc \citep{2010ApJ...714..927A}. While their corresponding $\gamma$-ray efficiencies are $\eta_{\gamma}$=$L_{\rm 1 GeV-1 TeV}$/$\dot{E}$ $\approx1.1\%$ and $1.6\%$, which are also comparable with the typical PWNe efficiencies observed at GeV energies \citep{2013ApJ...773...77A}.
In addition, the $\gamma$-rays of HESS J1813-178 show an energy-dependent morphology with its extension decreasing from GeV to TeV energies, which is typical for PWNe \citep{2020A&A...640A..76P,2024ApJ...961..213G}.
%The X-ray and the TeV luminosities are calculated to be $L_{\rm 0.3-10 keV}\approx 5.4\times 10^{34}~(d/12.0~\rm{kpc})^2~erg~s^{-1}$, $L_{\rm 1-10 TeV}\approx 1.4\times 10^{35}~(d/12.0~\rm{kpc})^2~erg~s^{-1}$, respectively. 

Adopting the updated $\gamma$-ray spectrum and the new X-ray observations for PWN, we first used a one-zone leptonic model to fit the multi-wavelength data of HESS J1813-178.
The radio upper limits toward the PWN region by VLA was also considered \citep{2007A&A...470..249F}.
For the ICS process in the leptonic model, the radiation field includes the cosmic microwave background (CMB), the infrared (IR) component from interstellar dust and gas with the typical values of T = 30K \& u = 1 eV $\rm cm^{-3}$, and the optical (OPT) component from stellar sources with T = 3000K \& u = 1 eV $\rm cm^{-3}$ \citep{2006ApJ...648L..29P, 2008ApJ...682..400P}.
For the leptonic model, the spectra of electrons was assumed to be a broken power-law model with an exponential cutoff (BPL) for typical PWNe \citep{2011MNRAS.410..381B}, which follows:
\begin{equation}
    \frac{dN_{\rm e}}{dE} \propto exp\left(-\dfrac{E}{E_{\rm e,cut}}\right) 
    \begin{cases}
    \left(\dfrac{E}{E_0}\right)^{-\alpha_{e,1}} \qquad \qquad \qquad \qquad ;  E < E_{\rm e,break}  \\
    \left(\dfrac{E_{\rm e,break}}{E_0}\right)^{\alpha_{e,2}-\alpha_{e,1}} \left(\dfrac{E}{E_0}\right)^{-\alpha_{e,2}} \,; E \geq E_{\rm e,break}
    \end{cases}
\end{equation}
Here, $\alpha_{\rm e,1}$ and $\alpha_{\rm e,2}$ are the spectral indices below and above the break energy of $\rm E_{e,break}$. The cutoff energy of electrons is denoted by $\rm E_{e,cut}$.
The distance of HESS J1813-178 as a PWN is adopted to be 12.0 kpc \citep{2021ApJ...917...67C}.
The CO observations in Section \ref{CO-observation} give an average gas density of 300$d_{12}^{-1}$ cm$^{-3}$ toward HESS J1813-178. Such density value would produce a quite high flux of bremsstrahlung component, which is much higher than that of ICS component considering the same electronic distribution.
Therefore, to reduce the contribution from bremsstrahlung process, we also adopted a low density value of 1.0 cm$^{\rm -3}$ by assuming that the similarity of the pulsar distance to the molecular distance from the observer is a coincidence.
The {\em naima} package was used to fit the multi-wavelength data with the Markov Chain Monte Carlo (MCMC) algorithm \citep{2015ICRC...34..922Z}.

The best-fit parameters of one-zone leptonic models for PWN are shown in Table \ref{table:model-leptonic}, and the corresponding modeled SEDs are given in the left panel of Figure \ref{fig:leptonic}.
For the ICS-dominated model with the gas density of 1.0 cm$^{\rm -3}$, the break energy of electrons is fitted to be $\sim$140 GeV. The spectral indices below and above it are 1.12 and 2.78, respectively.
The cutoff energy of $\sim$700 TeV is obtained, with the total energy of electrons above 1 GeV of $\sim$10$^{\rm 49}$ erg.
A magnetic field strength of about 7.3 $\mu$G is needed to explain the flux in the X-ray band.
%The much lower upper limits in the radio band need a low-energy cutoff of $\sim$30 GeV for the electronic spectrum.
Nonetheless, the hard X-ray spectrum given by \citet{2020MNRAS.498.4396H} suggests a harder electronic spectrum above the break energy, which can not be in accord with the flat $\gamma$-ray spectrum.

With the gas density of 300 cm$^{\rm -3}$, the $\gamma$-ray spectrum of HESS J1813-178 is dominated by the bremsstrahlung process, as shown in the middle panel of Figure \ref{fig:leptonic}. 
And the spectral indices of electrons below and above the break energy of $\sim$125 GeV are fitted to be 1.98 and 2.33, together with the cutoff energy of $\sim$400 TeV.
The magnetic field strength is similar to that of the one-zone ICS-dominated model, while the total energy of electrons is about one magnitude lower than that.
Although the synchrotron component of the Brem-dominated model could better accord with the radio upper limits and the hard X-ray spectrum without a low-energy cutoff of electrons, the curved $\gamma$-ray spectrum produced by the bremsstrahlung process can not match the $\gamma$-ray observations well, particularly at the lowest energies.

Given the hard X-ray spectrum and the flat GeV $\gamma$-ray spectrum, we also considered a two-zone leptonic model: one zone (zone 1) contains the newly accelerated electrons with the highest energies, which produces the X-ray emission. Another zone (zone 2) contains the old electrons accumulated in the region of GeV $\gamma$-ray emission. Such two-zone leptonic model could also explain the smaller extension of X-ray emission compared with the GeV or TeV $\gamma$-ray emission. 
For the electronic distribution of each zone, a single power-law with an exponential cutoff (PL) model is adopted in the form of 
\begin{equation}
\frac{dN_{\rm e,i}}{dE} \propto \left(\dfrac{E}{E_0}\right)^{-\alpha_{e,i}} exp\left(-\dfrac{E}{E_{\rm e,cut,i}}\right)
\end{equation}
where $\alpha_{\rm e, i}$ and $\rm E_{e,cut,i}$ (i = 1 or 2) are the spectral index and the cutoff energy of electrons for each zone. 
The low density value of 1.0 cm$^{\rm -3}$ is adopted for this model.
The best-fit parameters of two-zone leptonic models for PWN are shown in Table \ref{table:model-leptonic}, with the modeled SED shown in the right panel of Figure \ref{fig:leptonic}.
For zone 1, the spectral index and cutoff energy of electrons are fitted to be 1.64 and $\sim$200 TeV to explain the hard X-ray and TeV $\gamma$-ray spectra. A high magnetic field strength of $\sim$8.5 $\mu$G is needed.
And the total energy of electrons for zone 1 is about 7.85$\times$10$^{\rm 46}$ erg.
The GeV $\gamma$-ray emission could be explained by the electrons in zone 2 with an index of 2.77. The fitted cutoff energy of electrons of $\sim$90 TeV is lower than that of zone 1, while total energy of electrons of  2.97$\times$10$^{\rm 50}$ erg in zone 2 is much higher than that of zone 1.
The magnetic field strength in zone 2 is much lower with $\sim$1.5 $\mu$G.
And the much lower upper limits in the radio band need a low-energy cutoff of $\sim$50 GeV for the electronic spectrum in zone 2.

The soft GeV spectrum of SrcA suggests that its $\gamma$-ray emission could be related to the escaped electrons from the PWN associated with PSR J1813-1749, like Vela X \citep{2011ApJ...743L...7H,2013ApJ...774..110G}.
The steep GeV spectrum of SrcA indicates an absence of $\gtrsim$100 GeV energy particles from the extended $\gamma$-ray emission region.
Considering a typical magnitude of magnetic field with $\sim\mu$G, the synchrotron cooling time-scale for particles emitting in
the GeV range is up to $\sim 10^{8}$ yr, which makes the synchrotron cooling effect to be implausible for the steep GeV spectrum of SrcA.
And the possible scenario is that the bulk
of the $\gtrsim$100 GeV particles have escaped from the GeV emission region into
the interstellar medium \citep{2011ApJ...743L...7H}.
The required diffusion coefficient is calculated to be D = r$^{\rm 2}$/4t $\simeq$ 2$\times$10$^{\rm 29}$ $\rm cm^{2} \ s^{-1}$, where the radius of GeV $\gamma$-ray emission region of SrcA is estimated to be r$\simeq$120pc, and the diffusion time is adopted to be the characteristic age of PSR J1813-1749 with t $\sim$ 5600 yr.
This value is similar to the Galactic diffusion coefficient, but much faster than Bohm diffusion for the magnetic field
of $\sim\mu$G in the $\gamma$-ray emission region of SrcA.

\begin{table}[!htb]
	\centering
	\caption {Parameters of the hadronic models}
	\resizebox{1.0\textwidth}{!}{
		\begin{tabular}{ccccccc}
			\hline \hline
			Component  & $\alpha_{\rm p}$ &  E$_{\rm p, cut}$(TeV)  & $\beta$  & W$_{\rm p}$(erg)   & $\chi^2/N_{\rm dof}$ \\
			\hline
			% 	      HESS J1813-178  &  $2.00^{+0.04}_{-0.05}$  &  $187.13^{+54.92}_{-44.90}$    &  0.5  &  $(6.51^{+1.01}_{-0.90})(\rm n_{gas} / 720 cm^{-3})^{-1} (\rm d / 4.8 kpc)^{2} \times 10^{47}$  & 66.23/38=1.74 \\
			%                  &  $2.10^{+0.03}_{-0.04}$  &  $601.68^{+156.05}_{-132.42}$  &  1.0  &  $(8.13^{+1.13}_{-1.13})(\rm n_{gas} / 720 cm^{-3})^{-1} (\rm d / 4.8 kpc)^{2} \times 10^{47}$  & 58.01/38=1.53 \\
			%                 &  $2.11^{+0.03}_{-0.03}$  &  $656.07^{+170.54}_{-143.70}$  &  2.0  &  $(8.33^{+1.20}_{-1.07})(\rm n_{gas} / 720 cm^{-3})^{-1} (\rm d / 4.8 kpc)^{2} \times 10^{47}$  & 59.33/38=1.56 \\
			HESS J1813-178  &  $2.07^{+0.03}_{-0.03}$  &  $261.17^{+68.46}_{-55.98}$    &  0.5  &  $(1.26^{+0.16}_{-0.15})(\rm n_{gas} / 300 cm^{-3})^{-1} (\rm d / 12.0 kpc)^{2} \times 10^{49}$  & 79.57/37=2.15 \\
			&  $2.14^{+0.02}_{-0.02}$  &  $737.18^{+143.90}_{-141.15}$  &  1.0  &  $(1.46^{+0.15}_{-0.15})(\rm n_{gas} / 300 cm^{-3})^{-1} (\rm d / 12.0 kpc)^{2} \times 10^{49}$  & 66.30/37=1.79 \\
			&  $2.15^{+0.02}_{-0.03}$  &  $727.46^{+160.65}_{-142.26}$  &  2.0  &  $(1.51^{+0.17}_{-0.16})(\rm n_{gas} / 300 cm^{-3})^{-1} (\rm d / 12.0 kpc)^{2} \times 10^{49}$  & 66.20/37=1.79 \\
			\hline
			%            SrcA  &  $2.85^{+0.11}_{-0.09}$  &  --  &  --  &  $(1.97^{+0.37}_{-0.30})(\rm n_{gas} / 260 cm^{-3})^{-1} (\rm d / 4.8 kpc)^{2} \times 10^{49}$  & 6.68/3=2.23 \\
			SrcA  &  $2.71^{+0.07}_{-0.06}$  &  --  &  --  &  $(3.76^{+0.43}_{-0.39})(\rm n_{gas} / 70 cm^{-3})^{-1} (\rm d / 12.0 kpc)^{2} \times 10^{50}$  & 10.09/4=2.52 \\
			\hline
			SNR G13.5+0.2  &  $2.58^{+0.09}_{-0.13}$  &  --  &  --  &  $(2.61^{+0.63}_{-0.60})(\rm n_{gas} / 10 cm^{-3})^{-1} (\rm d / 13.0 kpc)^{2} \times 10^{50}$  & 4.59/3=1.53 \\
			\hline \hline
	\end{tabular}}
	\tablecomments{The total energy of protons, $W_{\rm p}$, is calculated for $E_{\rm p} > 1$ GeV.}
	\label{table:model-hadronic}
\end{table}  

\begin{figure*}[!htb]
	\centering
	\includegraphics[width=0.5\textwidth]{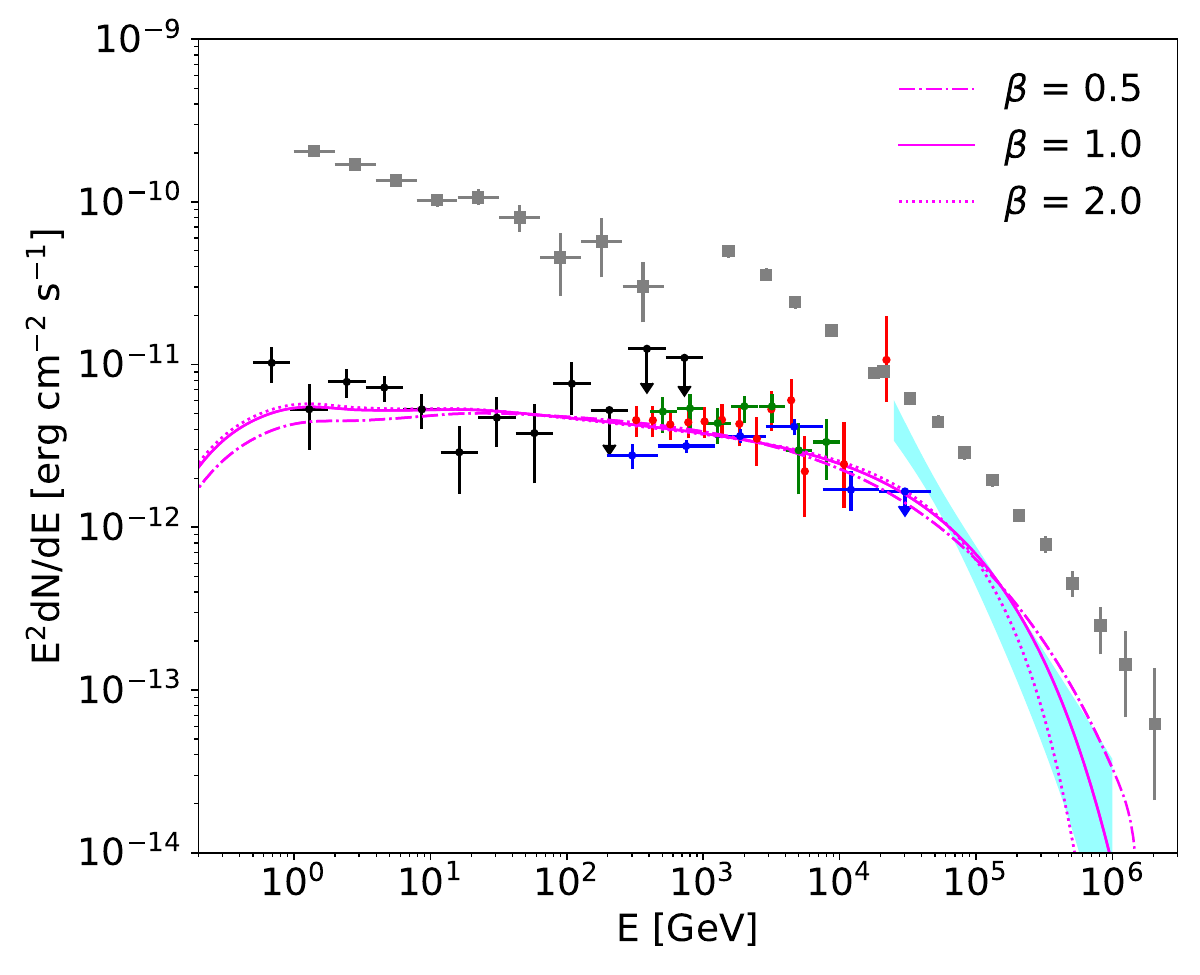}
	\caption{The hadronic model fitting for HESS J1813-178 with the different values of the sharpness of the cutoff, $\beta$. The gray square data show the $\gamma$-ray spectrum of Cygnus Cocoon detected by {\em Fermi}-LAT and LHAASO \citep{2017A&A...600A.107Y,2023arXiv231010100L}.}
	\label{fig:hadronic}
\end{figure*}

Now, the scenario that the $\gamma$-ray emission of HESS J1813-178 is associated with SNR G12.82-0.02 is considered.
The $\gamma$-ray SNRs could be roughly divided into two classes with  different spectra \citep{2015ARNPS..65..245F,2019ApJ...874...50Z}: the young-aged SNRs, including RX J1713.7-3946 \citep{2018A&A...612A...6H} and RX J0852.0-4622 \citep{2018A&A...612A...7H}, have relatively harder GeV $\gamma$-ray spectra, which are suggested to be from the leptonic process. The old-aged SNRs interacting with molecular clouds, like IC 443 and W44 \citep{2013Sci...339..807A}, have softer $\gamma$-ray spectra with a break at $\sim$ GeV, whose $\gamma$-ray emission are suggested to be from the hadronic process.
Considering the spatial association between the $\gamma$-ray emission from HESS J1813-178 and the molecular gas distribution, the hadronic model is adopted.
The spectrum of protons is assumed to be a single power-law with an exponential cutoff in the form of 
\begin{equation}
    \frac{dN_{\rm p}}{dE} \propto \left(\dfrac{E}{E_0}\right)^{-\alpha_{p}} exp\left(-\dfrac{E}{E_{\rm p,cut}}\right)^{\beta} 
\end{equation}
where $\alpha_{\rm p}$ and $\rm E_{p,cut}$ are the spectral index and the cutoff energy of protons, respectively. 
$\beta$ describes the sharpness of the cutoff, and the typical values of 0.5, 1.0, and 2.0 are adopted to constrain the parameters in the model.
The best-fit parameters of hadronic models for HESS J1813-178 are shown in Table \ref{table:model-hadronic}, and the corresponding modeled $\gamma$-ray spectrum is given in Figure \ref{fig:hadronic}.
The hadronic models with the three different values of $\beta$ can reproduce the $\gamma$-ray spectrum with little differences.
The spectral indices of protons are fitted to be about 2.1, while the cutoff energy varies from 261 TeV for $\beta$=0.5 to 737 TeV for $\beta$=1.0.
The total energy of protons above 1 GeV, W$_{\rm p}$, is estimated to be about (1.26-1.51) $\times$ 10$^{\rm 49}$ (n$_{\rm gas}$/300 cm$^{\rm -3}$)$^{-1}$ (d/12.0 kpc)$^{2}$ erg,
corresponding to $\sim$1\% particle acceleration efficiency for the explosion energy of a typical supernova (SN; E$_{\rm SN}$ $\sim$ $10^{51}$ erg).

\citet{2005ApJ...629L.105B} limited the age of SNR G12.82-0.02 from 285 to 2520 yrs with the assumptions of the distance of 4 kpc and the ambient density of 1.0 cm$^{\rm -3}$. 
Considering the far distance of 12 kpc and the high density associated with the molecular cloud around the remnant, the age of G12.82-0.02 should be much older. 
Such old-aged SNR would not expect to accelerate the needed protons with energy of hundreds of TeV \citep{2008ARA&A..46...89R}.
However, considering the much larger extension of the GeV $\gamma$-ray emission and the clouds compared with the radio size of SNR G12.82-0.02, cosmic rays that escaped from the SNR in earlier evolutional epochs
would be the possible explanation.
Such scenario is similar to a proposed origin of the $\gamma$-ray emission around SNR DA 530 \citep{2023ApJ...955...84X} and the SNR associated with PSR J0837-2454 \citep{2023ApJ...951..142Z}.
Given the typical value of Galactic diffusion coefficient, $\rm D(E)=3\times 10^{28} (E/10GeV)^{\delta}$ \citep[$\delta$ = 1/3 or 1/2 for Kolmogorve or Kraichnan turbulence;][]{2013A&ARv..21...70B}, and the radius of the GeV $\gamma$-ray emission region of $\rm r = 43\ pc$, the diffusion time for protons with energy of 100 GeV is calculated to be $\rm t_{diff} = r^2/4D(E)$ $\simeq$ 500-2000 yr for different values of $\delta$. 
Such time is still lower than the age of SNR G12.82-0.02, which is reasonable for such explanation. 
And the molecular
clouds can accumulate those cosmic rays in the past since the
diffusion coefficient is expected to be low in dense environments
\citep{1996A&A...309..917A}.
%However, the hard GeV $\gamma$-ray spectrum of HESS J1813-178 is much distinguished from the typical old-aged SNRs interacting with molecular clouds, which makes the SNR scenario still doubtable.

\subsubsection{Young Stellar Cluster scenario at the distance of 4.8 kpc}

Over the last decade, more and more observations support the young stellar clusters associated with star forming regions to be an important class of factories of Galactic cosmic rays \citep{2019NatAs...3..561A}.
Cosmic rays could be efficiently accelerated by strong fast winds of young massive stars and shocks caused by the core-collapse SNs in clusters.
The acceleration efficiency and maximum energy of particles are expected to be enhanced with respect to the
standard values derived for single SNR shock, considering the fact that the SN blast wave could interact with fast stellar winds \citep{2020SSRv..216...42B}.
Meanwhile, YSCs typically host the dense molecular gas to drive the strong star formation, which makes the hadronic interaction of accelerated CRs with the surrounding dense gas a natural explanation for their $\gamma$-ray emission.
So far, several such systems, e.g. Cygnus cocoon associated with the compact cluster Cygnus OB2 \citep{2011Sci...334.1103A, 2019NatAs...3..561A}, Westerlund 1 \citep{2012A&A...537A.114A}, Westerlund 2 \citep{2018A&A...611A..77Y}, NGC 3603 \citep{2017A&A...600A.107Y, 2020ApJ...897..131S}, and 30 Dor C \citep{2015Sci...347..406H}, have been detected with the extended $\gamma$-ray structures from GeV to TeV bands.
Especially, LHAASO recently detected the ultra high energy (UHE; $>$100TeV) $\gamma$-ray emission toward Cygnus cocoon in the energy range extending to few PeV \citep{2023arXiv231010100L}.
The $\gamma$-ray spectra of YSCs are relatively hard with the typical index of 2.1-2.3 \citep{2019NatAs...3..561A}.

The association between the $\gamma$-ray emission from HESS J1813-178 and YSC Cl 1813-178 has been discussed in the previous works \citep{2011ApJ...733...41M,2018ApJ...859...69A}.
With the updated GeV $\gamma$-ray spectrum, the YSC scenario with the hadronic model is also considered here.
The fitting results shown in Figure \ref{fig:hadronic} is same to that of the SNR scenario. 
However, for the YSC scenario, the total energy of protons above 1 GeV, W$_{\rm p}$, is calculated to be (0.84-1.01) $\times$ 10$^{\rm 48}$ (n$_{\rm gas}$/720 cm$^{\rm -3}$)$^{-1}$ (d/4.8 kpc)$^{2}$ erg.
Here, the gas density of 720 cm$^{\rm -3}$ calculated with the distance of 4.8 kpc for Cl 1813-178 is adopted.
The spectral index of proton distribution is similar to that of the typical YSCs, while the total energy for HESS J1813-178/Cl 1813-178 is about one order of magnitude lower than that of other YSCs \citep{2019NatAs...3..561A}.

The near-infrared spectroscopic survey in the direction of Cl 1813-178 cluster confirmed 25 massive stars, including 2 Wolf-Rayet (WR) stars, a candidate luminous blue variable (cLBV) and 21 OB stars  \citep{2011ApJ...733...41M}.
Different from the other known LBVs in clusters, which are typically among the most luminous members, the cLBV in Cl 1813-178 appears to be rather faint and has a mass smaller than that of the two detected WR stars.
Given the typical wind powers expected for a 35 M$_{\odot}$ star on and off the main sequence and in the WR stage \citep{2004A&A...424..747P} and ignoring the contribution from the cLBV, we get the total kinetic wind power of Cl 1813-178 with $\sim 10^{\rm 38} \rm erg \ s^{-1}$.
The kinetic power is similar to that of Cygnus Cocoon with 2$\times$10$^{\rm 38}$ $\rm erg \ s^{-1}$\citep{2011Sci...334.1103A}, but much lower than that of Westerlund 1 with 1$\times$10$^{\rm 39}$ $\rm erg \ s^{-1}$ \citep{2019NatAs...3..561A} and Westerlund 2 with 2$\times$10$^{\rm 40}$ $\rm erg \ s^{-1}$ \citep{2007A&A...463..981R}.
With the distance of 4.8 kpc, the GeV $\gamma$-ray luminosity of HESS J1813-178 is calculated to be about 9.6$\times$10$^{\rm 34}$ (d/4.8 kpc)$^{2}$ $\rm erg \ s^{-1}$, which represents
$\sim$0.1\% of the stellar wind power
in Cl 1813-178.
The conversion efficiency is slightly higher than that of Cygnus Cocoon with $\sim$0.03\% \citep{2011Sci...334.1103A}, Westerlund 1 with $\sim$0.002-0.02\% \citep{2023A&A...671A...4H} and Westerlund 2 with $\sim$0.005\% \citep{2018A&A...611A..77Y}.
However, the GeV extension of HESS J1813-178 with r = 17 pc for d = 4.8 kpc is much smaller that that of other YSCs \citep[e.g. r = 50 pc for Cygnus Cocoon, r = 60 pc for Westerlund 1, and r = 210 pc for Westerlund 2;][]{2019NatAs...3..561A,2018A&A...611A..77Y}.
If the GeV extension of r = 17 pc reflects the propagation depth
of CRs, the diffusion coefficient is calculated to be D = r$^{\rm 2}$/4T $\simeq$ 5.5$\times$10$^{\rm 24}$ $\rm cm^{2} \ s^{-1}$, where the age of Cl 1813-178 cluster is adopted to be T = 4$\times$10$^{\rm 6}$ yr \citep{2011ApJ...733...41M}.
While for the interstellar magnetic field B $\sim$ 10 $\mu$G, the Bohm diffusion coefficient is $\rm D_{B}$ = $\rm R_{L}c/3$ $\simeq$ $\rm 3\times10^{24}(E/1TeV)(B/10\mu G)^{-1}$ $\rm cm^{2} \ s^{-1}$.
And for the 1 TeV protons responsible for 100 GeV $\gamma$-rays, the diffusion coefficient in HESS J1813-178 is close to Bohm limit, which seems unrealistic.
Such result implies that the radius of the CR halo around
Cl 1813-178 is most likely significantly larger than that of the $\gamma$-ray emission region, like Westerlund 1 \citep{2019NatAs...3..561A}.

Considering the molecular gas around SrcA, we also used the hadronic model to fit the $\gamma$-ray spectrum of SrcA, by assuming a power-law distribution for protons.
The modeled spectrum is shown in Figure \ref{fig1:sed}, together with the predicted $\gamma$-ray emission assuming that the CR spectra therein are the same as those measured locally by AMS-02 \citep{2015PhRvL.114q1103A}.
The best-fit parameters of hadronic models are listed in Table \ref{table:model-hadronic}.
For SrcA, the spectral of protons with an index of $\sim$ 2.71 is need to explain its soft $\gamma$-ray spectrum detected.
With the different values of gas density derived from the different distances for SrcA, 
the total energy of protons above 1 GeV is estimated to be $W_{\rm p} \sim 3.76 \times 10^{50}(n_{\rm gas}/70\ {\rm cm}^{-3})^{-2}(d/12.0\ {\rm kpc})^2$ erg or $\sim 2.4 \times 10^{49}(n_{\rm gas}/170\ {\rm cm}^{-3})^{-2}(d/4.8\ {\rm kpc})^2$ erg.
The extended $\gamma$-ray morphology and the association with the molecular gas for SrcA suggest that the high energy protons could be accelerated and escaped from YSC Cl 1813-178 associated with HESS J1813-178.
Because of the energy-dependence of the diffusion coefficient, the initial particle spectrum of $dN/dE \propto E^{-\alpha}$ would be modified to be $dN/dE \propto E^{-(\alpha+\delta)}$ for the escaped particle spectrum with the continuous injection \citep{2019NatAs...3..561A}.
With the values of $\alpha \sim 2.1$ and $\delta=1/2$, the escaped proton spectrum follows $dN/dE \propto E^{-2.6}$, which is close to the fitted proton spectrum above.
The radius of the $\gamma$-ray emission region for SrcA is about 47 pc with the distance of 4.8 kpc.
If the escaping scenario between SrcA and YSC Cl 1813-178 is real, the diffusion length of 47 pc constrains that the diffusion coefficient should be as low as 10$^{\rm 25}$ $\rm cm^{2} \ s^{-1}$ by equating the diffusion time to be the age of cluster, which is also close to Bohm limit.

SrcC is spatially coincident with SNR G13.5+0.2. And the hadronic model is also used to fit its $\gamma$-ray spectrum.
By adopting the distance of 13.0 kpc \citep{2020AJ....160..263L} and assuming the gas density of 10 cm$^{\rm -3}$, 
the total energy of protons for SNR G13.5+0.2 is calculated to be $\sim 2.61 \times 10^{50}(n_{\rm gas}/10\ {\rm cm}^{-3})^{-2}(d/13.0\ {\rm kpc})^2$ erg, which means that $\sim$26\% of the SN kinetic energy of $\sim$10$^{\rm 51}$ erg is transferred to the energy of particles.

\section{Conclusions}
\label{con}

In this work, we analyzed the GeV $\gamma$-ray emission in the field of HESS J1813-178, using 14 years of {\em Fermi}-LAT data, 
and found that the GeV emission around HESS J1813-178 is composed of three $\gamma$-ray components: SrcA, SrcB and SrcC.
SrcC is spatially coincident with SNR G13.5+0.2.
SrcB has a small extension, which is spatially coincident with HESS J1813-178.
The GeV $\gamma$-ray spectrum follows a power-law model with an index of $2.11 \pm 0.08$, which can connect with the TeV SED of HESS J1813-178 smoothly.
The spatial and spectral associations indicate that SrcB could be the GeV counterpart of HESS J1813-178.
SrcA with a large size extension has a soft GeV $\gamma$-ray spectrum described by a log-parabola model.
No clear identified counterparts in other wavelengths are found in the region of SrcA.
While the CO observations display that the intensity of the $\gamma$-ray emission from SrcA and HESS J1813-178 spatially correlate with the molecular gas distribution in the velocity range of 45-60 km s$^{\rm -1}$.
The velocity range indicates two candidate kinematic distances: the near one is in accord with the distance of the young stellar cluster, Cl 1813-178, of $\sim$4.8 kpc, and the far one is in accord with the distance to the PSR J1813-1749/SNR G12.82-0.02 system, of $\sim$12 kpc.

We then discussed the possible $\gamma$-ray origin of HESS J1813-178 with three scenarios: the PWN associated with PSR J1813-1749, SNR G12.82-0.02 or YSC Cl 1813-178.
For the one-zone leptonic scenario with PWN, the electron distribution is adopted to be a broken power-law model with an exponential cutoff. 
The one-zone ICS-dominated and bremsstrahlung-dominated models with the different values of gas density are considered. 
However, none of them can well describe the multi-wavelength data from radio, X-ray to $\gamma$-ray bands. 
The one-zone ICS-dominated model expects a flat X-ray spectrum, which can not match the observed one with an spectral index of 1.5 $\pm$ 0.1 in the energy range of 2-10 keV.
With the high density around HESS J1813-178, the one-zone bremsstrahlung-dominated model could better accord with the hard X-ray spectrum.
While it predicts a curved $\gamma$-ray spectrum dominated by the bremsstrahlung process, which can not well match the $\gamma$-ray observations.
A two-zone leptonic model for PWN could explain the multi-wavelength data. The electron distribution of each zone is adopted to be a single power-law with an exponential cutoff. 
In one zone, an electron distribution with an index of $\sim$1.64 could explain the hard X-ray and TeV $\gamma$-ray spectra, with a high magnetic field strength of $\sim$8.5$\mu$G. In another zone, an electron distribution with an index of $\sim$2.77 and a low-energy cutoff of $\sim$50 GeV is use to explain the radio upper limits and the GeV $\gamma$-ray emission, with a low magnetic field strength of $\sim$1.5$\mu$G.
 
For the SNR scenario, only the hadronic model is discussed, considering the association between the $\gamma$-ray emission of HESS J1813-178 and the surrounding molecular gas.
The spectra of protons was assumed to be a single power-law with an exponential cutoff, and the different values of the sharpness of the cutoff, $\beta$, are also adopted.
The hadronic models with the different values of $\beta$ can reproduce the $\gamma$-ray spectra with little differences, and the proton distribution needs to be hard with an index of $\sim$2.1.
The cutoff energy varies between 261 TeV and 737 TeV for the different values of $\beta$.
The large extension of the GeV $\gamma$-ray emission and the clouds compared with the size of SNR G12.82-0.02 suggests that these protons contributed to the $\gamma$-ray emission could escape from the SNR in earlier evolutional epochs. The calculated diffusion time is lower than the age of SNR G12.82-0.02, which is reasonable for such explanation.
%Nonetheless, the hard GeV $\gamma$-ray spectrum of HESS J1813-178 is much distinguished from that of the typical old-aged SNRs interacting with molecular clouds, which typically have soft $\gamma$-ray spectra with a break at $\sim$GeV.
%Therefore, the SNR scenario as the $\gamma$-ray origin of HESS J1813-178 is still doubtable.

Although the flat $\gamma$-ray spectrum of HESS J1813-178 is similar to that of YSCs with the typical spectral index of 2.1-2.3,
%which supports the YSC scenario for the $\gamma$-ray emission of HESS J1813-178.
the GeV extension of HESS J1813-178 is much smaller than that of other YSCs.
Considering the GeV $\gamma$-ray size as the propagation depth of CRs, the calculated diffusion coefficient needs to be close to Bohm limit, which is not very reasonable.

The soft $\gamma$-ray spectrum of SrcA can be explained by the fast diffusion of electrons from the PWN associated with PSR J1813-1749, like Vela X, or the hadronic process due to the inelastic collisions between the molecular cloud and the high energy protons, which can be accelerated in and escaped from YSC Cl 1813-178.

It should be noted that the hadronic models show that the highest energy of protons should exceed a few hundred TeV, which makes HESS J1813-178 to be a promising PeV CR accelerator (PeVatron). 
LHAASO has detected the UHE $\gamma$-ray around HESS J1813-178, and more detailed analysis based on more LHAASO data, especially the morphological analysis, would be helpful to investigate the origin of the $\gamma$-ray emission in this region and test its PeVatron nature.
Moreover, the sensitivity of LHAASO above 10 TeV could help to study in
detail the shape of the particle spectrum cutoff, that encodes
information about the particle acceleration and transport mechanisms \citep{2007A&A...465..695Z, 2017APh....88...38R}.

\vspace{0.5cm}
%\begin{acknowledgments}
We would like to thank the anonymous referee for very helpful comments, which help to improve the paper. 
This work is based on observations made with NASA's Fermi Gamma-Ray Space Telescope, and it also makes use of molecular line data from FOREST Unbiased Galactic plane Imaging survey with the Nobeyama 45 m telescope (FUGIN). This work is supported by the National Natural Science Foundation of China under the grants 12103040.
%\end{acknowledgments}


\bibliography{sample631}{}
\bibliographystyle{aasjournal}



\end{document}